\documentclass[aps,prl,reprint,superscriptaddress,showpacs,longbibliography,footnoteinbib]{revtex4-1}
\usepackage[latin9]{inputenc}
\usepackage{xcolor}
\usepackage{pdfcolmk}
\usepackage{float}
\usepackage{bm}
\usepackage{amsmath}
\usepackage{amssymb}
\usepackage{graphicx}
\PassOptionsToPackage{normalem}{ulem}
\usepackage{ulem}
\usepackage[unicode=true,
 bookmarks=false,
 breaklinks=false,pdfborder={0 0 1},backref=false,colorlinks=true]
 {hyperref}
\hypersetup{pdftitle={Title},
 plainpages=false,pdfpagelabels,linkcolor=blue,urlcolor=blue,citecolor=blue,pdfdisplaydoctitle=true,pdfduplex=DuplexFlipLongEdge}

\makeatletter

\providecommand{\tabularnewline}{\\}
\providecolor{lyxadded}{rgb}{0,0,1}
\providecolor{lyxdeleted}{rgb}{1,0,0}
\DeclareRobustCommand{\lyxsout}[1]{\ifx\\#1\else\sout{#1}\fi}

\usepackage{units}\usepackage{wasysym}

\definecolor{orange}{rgb}{0.50, 0.20, 0.0}

\newcommand{\beginsupplement}{%
	\setcounter{page}{1}
	 \renewcommand{\thepage}{SM - \arabic{page}}%
        \setcounter{table}{0}
        \renewcommand{\thetable}{S\arabic{table}}%
        \setcounter{figure}{0}
        \renewcommand{\thefigure}{S\arabic{figure}}%
        \setcounter{section}{0}
        \renewcommand{\thesection}{S\arabic{section}}%
        \setcounter{section}{0}
        \renewcommand{\thesection}{S\arabic{section}}%
        \setcounter{subsection}{0}
        \renewcommand{\thesubsection}{S\arabic{section}.\arabic{subsection}}%
        \setcounter{equation}{0}
        \renewcommand{\theequation}{S\arabic{equation}}%

     }

\usepackage{bm}
\usepackage{braket}

\makeatother

\begin{document}
\noindent\begin{minipage}[t]{1\columnwidth}%
\global\long\def\ket#1{\left| #1\right\rangle }%

\global\long\def\bra#1{\left\langle #1 \right|}%

\global\long\def\kket#1{\left\Vert #1\right\rangle }%

\global\long\def\bbra#1{\left\langle #1\right\Vert }%

\global\long\def\braket#1#2{\left\langle #1\right. \left| #2 \right\rangle }%

\global\long\def\bbrakket#1#2{\left\langle #1\right. \left\Vert #2\right\rangle }%

\global\long\def\av#1{\left\langle #1 \right\rangle }%

\global\long\def\tr{\text{tr}}%

\global\long\def\Tr{\text{Tr}}%

\global\long\def\pd{\partial}%

\global\long\def\im{\text{Im}}%

\global\long\def\re{\text{Re}}%

\global\long\def\sgn{\text{sgn}}%

\global\long\def\Det{\text{Det}}%

\global\long\def\abs#1{\left|#1\right|}%

\global\long\def\up{\uparrow}%

\global\long\def\down{\downarrow}%

\global\long\def\vc#1{\mathbf{#1}}%

\global\long\def\bs#1{\boldsymbol{#1}}%

\global\long\def\t#1{\text{#1}}%
\end{minipage}
\title{Incommensurability enabled quasi-fractal order in 1D narrow-band moiré
systems}
\author{Miguel Gonçalves}
\affiliation{CeFEMA-LaPMET, Departamento de Física, Instituto Superior Técnico,
Universidade de Lisboa, Av. Rovisco Pais, 1049-001 Lisboa, Portugal}
\author{Bruno Amorim}
\affiliation{Centro de Física das Universidades do Minho e do Porto (CF-UM-UP),
Laboratório de Física para Materiais e Tecnologias Emergentes (LaPMET),
Universidade do Minho, Campus de Gualtar, 4710-057 Braga, Portugal}
\author{Flavio Riche}
\affiliation{CeFEMA-LaPMET, Departamento de Física, Instituto Superior Técnico,
Universidade de Lisboa, Av. Rovisco Pais, 1049-001 Lisboa, Portugal}
\author{Eduardo V. Castro}
\affiliation{Centro de Física das Universidades do Minho e Porto, Departamento
de Física e Astronomia, Faculdade de Ciências, Universidade do Porto,
4169-007 Porto, Portugal}
\affiliation{Beijing Computational Science Research Center, Beijing 100193, China}
\author{Pedro Ribeiro}
\affiliation{CeFEMA-LaPMET, Departamento de Física, Instituto Superior Técnico,
Universidade de Lisboa, Av. Rovisco Pais, 1049-001 Lisboa, Portugal}
\affiliation{Beijing Computational Science Research Center, Beijing 100193, China}
\begin{abstract}
We demonstrate that quasiperiodicity can radically change the ground
state properties of 1D moiré systems with respect to their periodic
counterparts. By studying an illustrative example we show that while
narrow bands play a significant role in enhancing interactions both
for commensurate and incommensurate structures, only quasiperiodicity
is able to extend the ordered phase down to an infinitesimal interaction
strength. In this regime, the quasiperiodic-enabled state has contributions
from infinitely many wave vectors. This quasi-factal regime cannot
be stabilized in the commensurate case even in the presence of a narrow
band. These findings suggest that quasiperiodicity may be a critical
factor in stabilizing non-trivial ordered phases in interacting moiré
structures and signal out multifractal non-interacting phases, recently
found in 2D incommensurate moiré systems, as particularly promising
parent states.
\end{abstract}
\maketitle
Quasiperiodic structures (QPS) exhibit a plethora of intriguing features,
such as non-trivial localization \citep{AubryAndre,Roati2008,Lahini2009,Schreiber842,Luschen2018,slager1,Huang2016a,PhysRevLett.120.207604,Park2018,PhysRevB.100.144202,Fu2020,Wang2020,Goncalves_2022_2DMat,PhysRevX.7.041047,PhysRevLett.123.070405,PhysRevLett.125.060401,PhysRevLett.126.110401},
multifractality \citep{PhysRevLett.110.146404,Liu2015,PhysRevB.93.104504,CadeZ2019,PhysRevLett.123.025301,PhysRevLett.125.073204,anomScipost},
and topological \citep{Kraus2012,PhysRevLett.109.116404,Verbin2013,PhysRevResearch.1.033009,Zilberberg:21,slager2,PhysRevResearch.3.013265}
properties, that can be encountered in diverse settings, including
optical lattices \citep{PhysRevA.75.063404,Roati2008,Modugno_2009,Schreiber842,Luschen2018,PhysRevLett.123.070405,PhysRevLett.125.060401,PhysRevLett.126.110401,PhysRevLett.126.040603,PhysRevLett.122.170403},
photonics, \citep{Lahini2009,Kraus2012,Verbin2013,PhysRevB.91.064201,Wang2020,https://doi.org/10.1002/adom.202001170}
phononics \citep{PhysRevLett.122.095501,Ni2019,PhysRevLett.125.224301,PhysRevApplied.13.014023,PhysRevX.11.011016,doi:10.1063/5.0013528},
and two-dimensional (2D) moiré materials \citep{Balents2020,Andrei2020}.

Another notable feature of QPS is the formation of narrow (nearly
flat) bands \citep{Balents2020,Andrei2020} in the energy spectrum
which enhance correlations in the presence of electron-electron interactions.
This enhancement mechanism is believed to underlie the abundance of
strongly-correlated phases recently observed in moiré materials.

Beyond its impact on the energy spectrum, quasiperiodicity also affects
eigenstates. Nevertheless, beyond localization effects, the impact
of incommensurability on the eigenstate structure has been traditionally
disregarded to explain the physics of moiré materials \citep{Balents2020,Andrei2020}.
However, recent research on twisted bilayer graphene (tBLG) and related
2D models \citep{Goncalves_2022_2DMat,Fu2020,PhysRevB.101.235121}
suggests that in the narrow-band regime incommensurability may induce
single-particle states to delocalize both in momentum and position
space, resulting in sub-ballistic transport. This suggests that the
traditional starting point to include interactions within the narrow-band
regime that assumes plane wave single-particle states must be revised.

To establish whether quasiperiodicity plays a role in the emergence
of correlated phases beyond the creation of narrow bands requires
numerically unbiased methods able to take electron-electron interactions
into account in large-scale QPS. In 2D, this is notably challenging
due to the lack of efficient numerical tools. In 1D, on the other
hand, ground-state properties can be efficiently studied for fairly
large system sizes using the density-matrix-renormalization-group
(DMRG) technique \citep{PhysRevLett.69.2863,RevModPhys.77.259}.

Although somehow less rich than their 2D counterparts \citep{PhysRevB.100.144202,Szabo2020,Fu2020,goncalves2022topological},
1D QPS may also host phases that are neither localized nor ballistic.
These so-called critical multifractal states were first identified
at localization-ballistic transitions and later shown to exist in
extended regions of the phase space. Interestingly, both in 1D quasiperiodic
critical phases and in the incommensurate-driven regime found in the
narrow band of tBLG's, eigenstates are delocalized in both real and
momentum space \citep{PhysRevLett.110.146404,Liu2015,PhysRevB.93.104504,PhysRevLett.123.025301,PhysRevLett.125.073204,CadeZ2019,anomScipost}.
This suggests that the study of interactions in the 1D setting may
provide relevant insights into 2D moiré structures.

With notable exceptions, most studies of interacting 1D QPS focus
on infinite-temperature physics and many-body localization phenomena
\citep{PhysRevB.87.134202,PhysRevA.92.041601,PhysRevLett.115.230401,PhysRevB.96.075146,Znidari,PhysRevResearch.1.032039,PhysRevB.100.104203,vu2021fermionic,PhysRevB.104.214201}.
Interestingly, even at infinite temperature, critical multifractal
states, emerging out of their non-interacting counterparts, have been
recently proposed \citep{PhysRevLett.126.080602}.

Only a few studies address ground-state localization properties \citep{PhysRevB.65.115114,PhysRevB.89.161106,SciPostPhys.1.1.010,PhysRevB.101.174203,PhysRevLett.126.036803,oliveira2023incommensurabilityinduced,goncalves2023shortrange}
. Surprisingly, some of these works revealed that interactions may
be irrelevant at criticality \citep{PhysRevB.101.174203,goncalves2023shortrange}.
However, the stability of extended multifractal critical phases to
interactions has not yet been addressed to our knowledge. Furthermore,
a comparison of how interactions differ in their effects on phases
of commensurate and incommensurate 1D moiré structures is also absent.
Understanding these effects from a theoretical standpoint in 1D QPS
could shed light on the rich physics of 2D moiré materials, for which
correlated phenomena was recently experimentally reported in strongly
quasiperiodic regimes \citep{uri2023superconductivity}. Experimentally
testing these findings would also open new research directions in
1D optical lattices where moire structures are routinely implemented.

\begin{figure}[h]
\centering{}
\includegraphics[width=1.0\columnwidth]{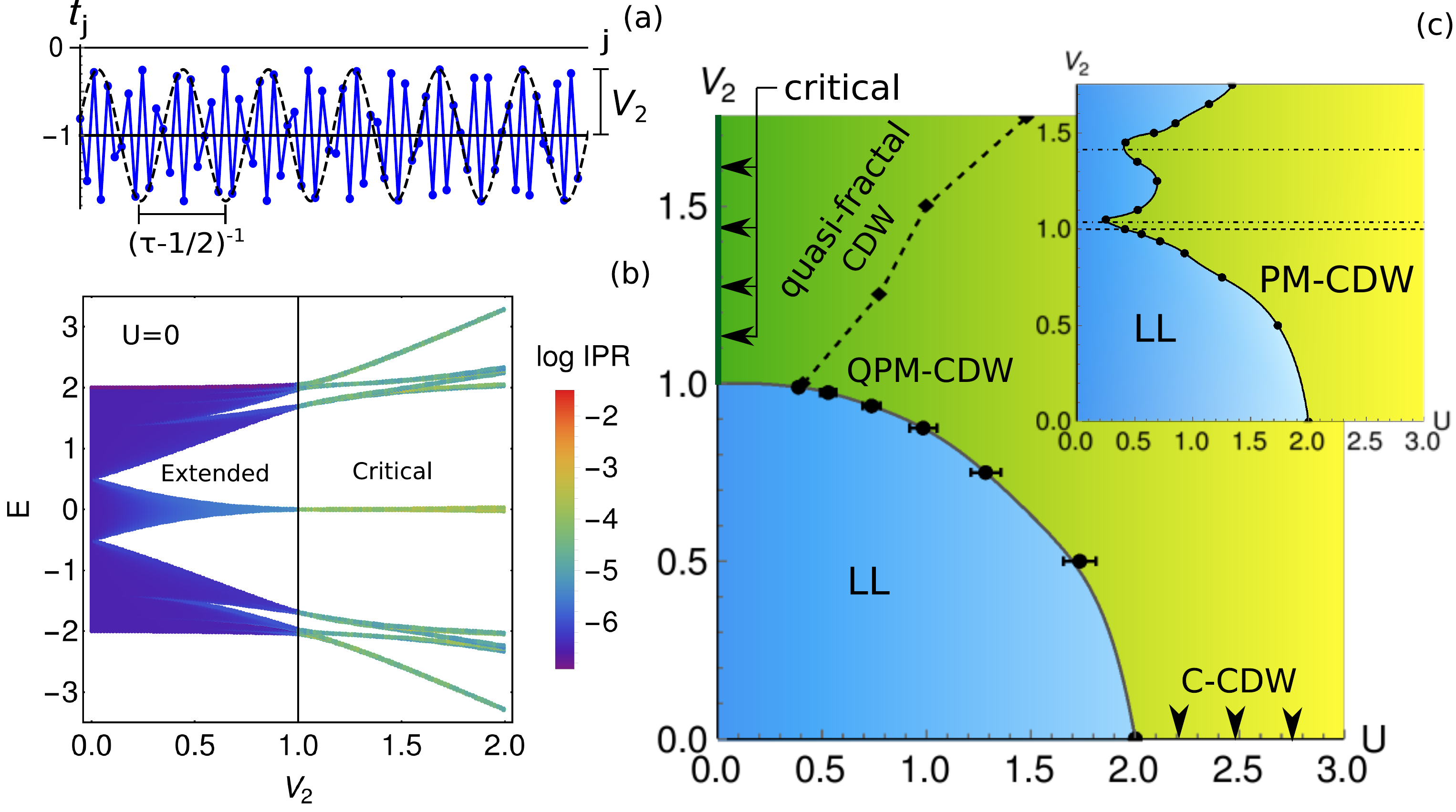}\caption{(a) Hopping modulation for the studied model (see below Eq.$\,$\ref{eq:Hamiltonian}),
with $\tau=0.5812$. This choice creates a moiré pattern of size $L_{M}=(\tau-1/2)^{-1}\approx12.3$.
(b) Single-particle spectrum of the model in the non-interacting limit
($U=0$) as function of $V_{2}$, for $\tau=587/1010$ and $N=1010$
sites, using periodic boundary conditions and $\phi=0$. $V_{2}=1$
separates phases with extended and critical eigenstates. Each eigenstate
is colored according to its inverse participation ratio (IPR) value,
defined for a given eigenstate at energy $E_{n}$ as ${\rm IPR}(E_{n})=\Big(\sum_{i=0}^{N-1}|\psi_{i}^{(n)}|^{4}\Big)/\Big(\sum_{i=0}^{N-1}|\psi_{i}^{(n)}|^{2}\Big)^{2}$,
with $\psi_{i}^{(n)}$ being the amplitude of the $n$-th eigenstate
at site $i$. (c) Main figure: Phase diagram of the incommensurate
moiré system in the plane of interaction strength $U$ and intensity
of quasiperiodic hoppings $V_{2}$, with Luttinger liquid (LL), C-CDW
(commensurate CDW) and QPM-CDW (quasiperiodic Moiré CDW) phases. The
full and dashed lines connecting the data points are guides to the
eye corresponding respectively to a phase transition and a crossover
(see text for calculation details). Inset: Phase diagram of a commensurate
Moiré system defined by $\tau=\tau_{c}=7/12$ and $\phi=0$, whose
exact unit cell is precisely the Moiré pattern. In this case there
is a periodic Moiré CDW (PM-CDW) phase at finite $V_{2}$ and $U$.
The dashed line is at $V_{2}=1$ and the dot-dashed lines mark points
in which the narrow-band's width vanishes in the commensurate case.
The lines connecting the data points are guides to the eye. \label{fig:main}}
\end{figure}

In this work, we establish that incommensurability may radically change
the many-body ground state of a strongly interacting system with respect
to similar commensurate structures. Our illustrative example is a
1D tight-binding model that, without interactions, hosts an extended
phase of multifractal critical states concomitant with a narrow energy
band. Quasiperiodicity is introduced in the modulation of nearest-neighbor
hopping \citep{PhysRevB.91.014108} with a period close to $2$ lattice
sites. The ensuing moiré patterns are shown in Fig.$\,$\ref{fig:main}(a).
Fig.$\,$\ref{fig:main}(b) shows the single-particle band structure
of the non-interacting model. A phase transition between an extended
(ballistic) and a multifactal critical phase arises when the strength
of the quasiperiodic modulation, $V_{2}$, reaches $V_{2}=1$. The
narrow band centered around zero energy becomes flatter as the critical
phase is approached. To target the effect of interactions in the physics
of the narrow band, we study the ground-state phase diagram of this
model in the presence of nearest-neighbor repulsive interactions,
with intensity $U$, at half-filling, corresponding to a Fermi energy
in the middle of the narrow band.

Our main results are summarized in Fig.$\,$\ref{fig:main}. In the
absence of quasiperiodic modulation ($V_{2}=0$) there is a well-known
transition from a (gapless) Luttinger Liquid (LL) phase into a (gapped)
commensurate CDW (C-CDW) phase with ordering wave-vector $\kappa=\pi$
\citep{RevModPhys.83.1405}. For $0<V_{2}<1$, upon increasing the
quasiperiodic modulation, the differences between commensurate and
incommensurate systems are not significant up to $V_{2}\lesssim1$:
in both cases the LL gapless extended phase is robust up to a finite
value of the interaction strength, $U$. The critical value $U=U_{c}(V_{2})$,
beyond which the transition to a gapped CDW phase occurs, decreases
with increasing $V_{2}$. We dub this ordered phase, characterized
by CDW order with multiple wave vectors, a moiré CDW. Depending on
the nature of the system, these additional wave vectors are either
commensurate or incommensurate with the lattice. We refer to them
as periodic (PM-CDW) or quasiperiodic (QPM-CDW) moiré CDW, respectively.

For $V_{2}\geq1$, commensurate and incommensurate modulations differ
drastically once interactions are turned on. For commensurate structures,
a gapless LL phase still occurs for small but finite $U$, transitioning
into a gapped PM-CDW phase for larger $U$. Contrastingly, incommensurability
suppresses the LL phase for $V_{2}\geq1$ rendering the system unstable
towards a QPM-CDW phase for any finite $U$. We dub this remarkable
QPM-CDW regime arising for small enough values of $U$ and $V_{2}\geq1$
a quasi-fractal CDW due to the quasi-fractal structure of the charge
order in this state. As $U$ increases there is a crossover from the
quasi-fractal CDW regime into an ordered state similar to the QPM-CDW
that occurs for $V_{2}<1$. As we move away from the quasi-fractal
regime, the moiré CDW of the incommensurate and commensurate systems
becomes increasingly indistinguishable.

The remainder of the paper is organized as follows. After providing
details on the model and method, we introduce the different observables
tailored to investigate the nature of the CDW phase. We analyze these
quantities to infer the impact of interactions in extended and critical
states. Finally, we discuss the results and their implications. In
the supplemental material (SM) section \citep{SM}, we provide additional
results including a thorough finite scaling analysis that further
supports our conclusions.

\paragraph*{Model and Methods.---}

We consider a tight-binding chain of spinless fermions in which the
nearest-neighbor hoppings are modulated quasiperiodically \citep{PhysRevB.42.8282,PhysRevB.50.11365,PhysRevB.91.014108}.
In addition, we take a repulsive interaction of magnitude $U>0$ between
nearest neighbors' densities. The Hamiltonian reads

\begin{equation}
\begin{aligned}H= & \sum_{j}t_{j}c_{j}^{\dagger}c_{j+1}+\textrm{h.c.}+U\sum_{j}n_{j}n_{j+1}\end{aligned}
,\label{eq:Hamiltonian}
\end{equation}
where $c_{j}^{\dagger}$ creates a particle at site $j$, $t_{j}=-1-V_{2}\cos[2\pi\tau(j+1/2)+\phi]$
is the hopping from site $j+1$ to site $j$, which has quasiperiodic
modulation of strength $V_{2}$ and period $\tau^{-1}$. Since $j$
is an integer , if $\tau$ is irrational, the periodicity of the hopping
term is infinite. In the following we consider the half-filled case,
to ensure a gapped ordered state exists at $\kappa=\pi$ for $V_{2}=0$,
and open boundary conditions (OBC), that are suitable for the DMRG
algorithm (in the SM \citep{SM} we present further results for twisted
boundary conditions).

Numerical results were obtained with DMRG \citep{PhysRevLett.69.2863,RevModPhys.77.259},
as implemented in the iTensor library \citep{itensor,itensorpaper}.
We required the iTensor's truncation error to be less than $10^{-12}$
and only stopped the sweeping procedure once convergence requirements
are satisfied, up to a maximum of $500$ sweeps. In most of the calculations
we require the energy variance, $\Delta_{H}=\langle H^{2}\rangle-\langle H\rangle^{2}$,
to be below $10^{-6}$; the ground-state energy difference between
two sweeps to be below $\Delta E_{{\rm GS}}=10^{-7}$; and the difference
in the entanglement entropy at the middle bond between two sweeps
to be below $\Delta S_{{\rm GS}}=10^{-4}$ \footnote{For the largest used system sizes, in some cases we slightly relaxed
this criterium to $\Delta E_{{\rm GS}}<10^{-7}$,$\Delta_{H}<5\times10^{-6}$
and $\Delta S_{{\rm GS}}<5\times10^{-4}$.}.

For the finite-size scaling analysis, we approximate the irrational
$\tau$ by a rational number $\tau\simeq\tau_{p,N}=p/N$, and take
$N$ to be the size of the chain containing a single unit cell. We
then consider a sequence of approximates $\tau_{p,N}$ of increasing
size, $N$.

Throughout the paper we take $\tau=1/2+\delta\simeq1/2$. This choice
ensures the formation of a narrow-band at the center of the energy
spectrum, as we detail in the SM \citep{SM}. The period of the corresponding
moiré pattern, $L_{m}=\left(\tau-1/2\right)^{-1}=\delta^{-1}$, increases
as $\tau$ approaches $1/2$. At the same time the central narrow-band
in Fig.$\,$\ref{fig:main} becomes increasingly flat (see \citep{SM})
, rendering DMRG convergence progressively difficult. For these reasons,
we set $\tau\simeq0.5812$. Furthermore, to decrease finite-size effects
when capturing the CDW order for twisted boundary conditions (in the
SM \citep{SM}), we consider system sizes with an even number of sites.
Also, we do not draw approximants from an exact sequence of rational
convergents of $\tau$, as this would severely limit the available
system sizes. Instead, having in mind that physical properties should
not depend on slight variations in $\tau$, we choose rational approximants
that are very close to $\tau=1453/2500=0.5812$, which we took to
be the largest system size containing $N=2500$ sites. The particular
series of chosen approximants follows,
\begin{center}
\begin{tabular}{|c|c|c|c|c|c|c|c|c|}
\hline 
$N$ & 112 & 198 & 308 & 504 & 1010 & 1478 & 2008 & 2500\tabularnewline
\hline 
\hline 
$\tau_{p,N}$ & $\frac{65}{112}$ & $\frac{115}{198}$ & $\frac{179}{308}$ & $\frac{293}{504}$ & $\frac{587}{1010}$ & $\frac{859}{1478}$ & $\frac{1167}{2008}$ & $\frac{1453}{2500}$\tabularnewline
\hline 
\end{tabular}.
\par\end{center}

\noindent We compare the incommensurate (IS) and commensurate systems
(CS) where the repeating unit cell is taken to coincide with the moiré
pattern wavelength. Concretely, we considered $\tau=\tau_{c}=7/12$,
corresponding to a unit cell with $12$ sites. 

\begin{figure}[H]
\centering{}\includegraphics[width=1\columnwidth]{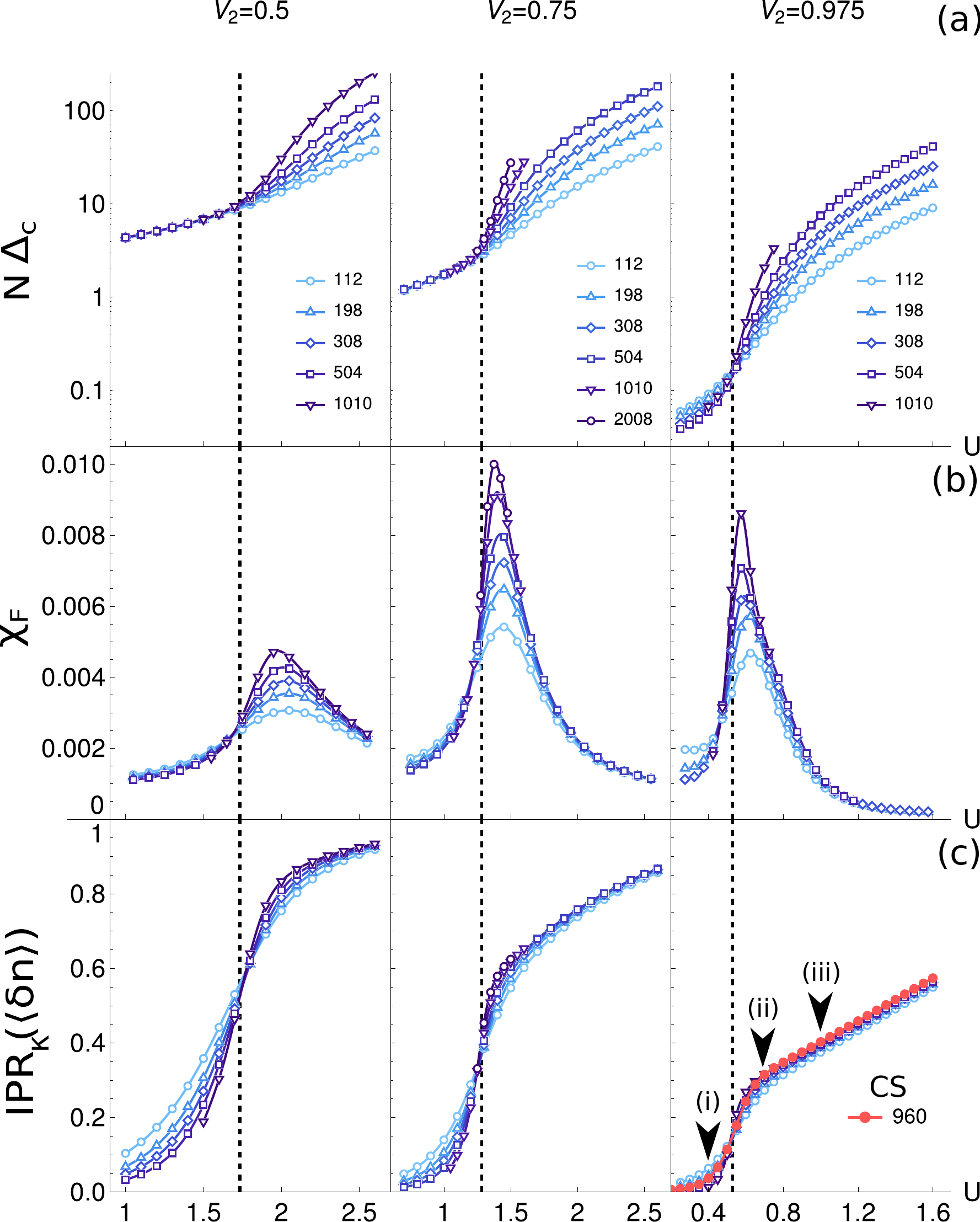}\caption{Finite-size scaling results for: (a) charge gap, $\Delta_{c}$, defined
in Eq.$\,$\ref{eq:cgap}; (b) normalized fidelity susceptibility,
$\chi_{F}$, defined in Eq.$\,$\ref{eq:chi_F}, computed for consecutive
$U=U_{i}$ in (a) and with $\delta U=U_{i+1}-U_{i}$; (c) ${\rm IPR}_{\kappa}(\langle\delta\bm{n}\rangle)$
defined in Eq.$\,$\ref{eq:iprk}. The top label indicates the value
of $V_{2}$ used in the corresponding column (a-c). The dashed lines
correspond to the thermodynamic limit extrapolation of the critical
point. In (c), results for the CS with $\tau_{c}=7/12$ are shown
in red, for $\phi=0$. All other results are for an IS with $\phi=\pi/4$
.\label{fig:LL-CDW}}
\end{figure}

For establishing the phase diagram we consider different quantities.
The first is the charge gap, which distinguishes between gapped and
gapless phases, and is defined as

\begin{equation}
\Delta_{c}=E(N_{p}+1)+E(N_{p}-1)-2E(N_{p})\label{eq:cgap}
\end{equation}
where $E(N_{p})$ is the ground-state energy for a filling of $N_{p}$
particles(at half-filling, $N_{p}=N/2$). In the (gapless) LL phase
$\lim_{N\rightarrow\infty}\Delta_{c}=0$, while $\Delta_{c}$ approaches
a non-zero constant within a gapped CDW phase. To unbiasedly detect
quantum phase transitions, we compute the fidelity between ground-states
at different values of $U$, defined as $F=|\braket{\Psi(U)}{\Psi(U+\delta U)}|$
\citep{fidelityRef}, where $\ket{\Psi(U)}$ denotes the many-body
ground-state for interaction strength $U$. More concretely, we study
the fidelity susceptibility normalized by the system size

\begin{equation}
\chi_{F}=\frac{-2\log F}{N\delta U^{2}},\label{eq:chi_F}
\end{equation}
in the limit $\delta U\rightarrow0$ \citep{fidelityRef}. At a quantum
phase transitions $\chi_{F}$ diverges as $N$ increases.

We also compute the average density fluctuations given by $\langle\delta n_{m}\rangle=\langle n_{m}\rangle-1/2$,
where the expectation value is taken with respect to the ground state,
and its discrete Fourier transform:

\begin{equation}
\langle\delta n_{\kappa}\rangle=\frac{1}{\sqrt{N}}\sum_{m=0}^{N-1}e^{{\rm i}\kappa m}\langle\delta n_{m}\rangle\label{eq:delta_nk}
\end{equation}
where $\kappa$ takes values $\kappa_{j}=2\pi j/N,\textrm{ }j=0,\cdots,N-1$,
which signals the formation of CDW phases. Since the model in Eq.$\,$\ref{eq:Hamiltonian}
is symmetric under the particle-hole transformation $c_{j}^{\dagger}\rightarrow(-1)^{j}c_{j}$
, we have that $\langle\delta n_{m}\rangle=0$ at half-filling. Therefore,
in order to detect ordered phases, we add an edge field term to the
Hamiltonian of Eq.~\eqref{eq:Hamiltonian} of the form $\epsilon c_{0}^{\dagger}c_{0}$,
setting $\epsilon=2$. In an ordered phase, the edge field ($\epsilon\neq0$)
selects a particular CDW state, inducing extensive density fluctuations
and therefore $\langle\delta n_{\kappa}\rangle\sim N^{1/2}$, for
the $\kappa$ wave-vectors contributing to the CDW charge fluctuations.
In a disordered phase, the edge field only induces non-extensive boundary
fluctuations and therefore $\langle\delta n_{\kappa}\rangle\sim N^{-1/2}$
at any $\kappa$. Thus a given wave-vector, $\kappa$, is present
in the CDW if $K_{\kappa}=\lim_{N\to\infty}\langle\delta n_{\kappa}\rangle/N^{1/2}$
is finite. To quantitatively distinguish between ordered and disordered
phases, we define an inverse participation ratio for the charge fluctuations
$\langle\delta n_{\kappa}\rangle$ as

\begin{equation}
{\rm IPR}_{\kappa}(\langle\delta\bm{n}\rangle)=\left(\sum_{\kappa=0}^{N-1}|\langle\delta n_{\kappa}\rangle|^{4}\right)/\left(\sum_{\kappa=0}^{N-1}|\langle\delta n_{\kappa}\rangle|^{2}\right)^{2}.\label{eq:iprk}
\end{equation}
In a disordered phase, there are only boundary fluctuations, implying
that ${\rm IPR}_{\kappa}(\langle\delta\bm{n}\rangle)\sim N^{-1}$;
while in a charge ordered phase there are finite bulk fluctuations
for any system size described by a non-extensive number of wave vectors
$\kappa$, implying that ${\rm IPR}_{\kappa}(\langle\delta\bm{n}\rangle)\sim N^{0}$.
For a conventional (commensurate) CDW phase with order only at $\kappa=\pi$,
we simply have ${\rm IPR}_{\kappa}(\langle\delta\bm{n}\rangle)=1$.
To improve the estimation of the phase transition points, we also
considered the entanglement entropy at the middle bond and the squared
fluctuations $\chi_{R}=N^{-1}\sum_{i}\langle\delta n_{i}\rangle^{2}$
(see SM \citep{SM}).

\begin{figure}[H]
\centering{}\includegraphics[width=1\columnwidth]{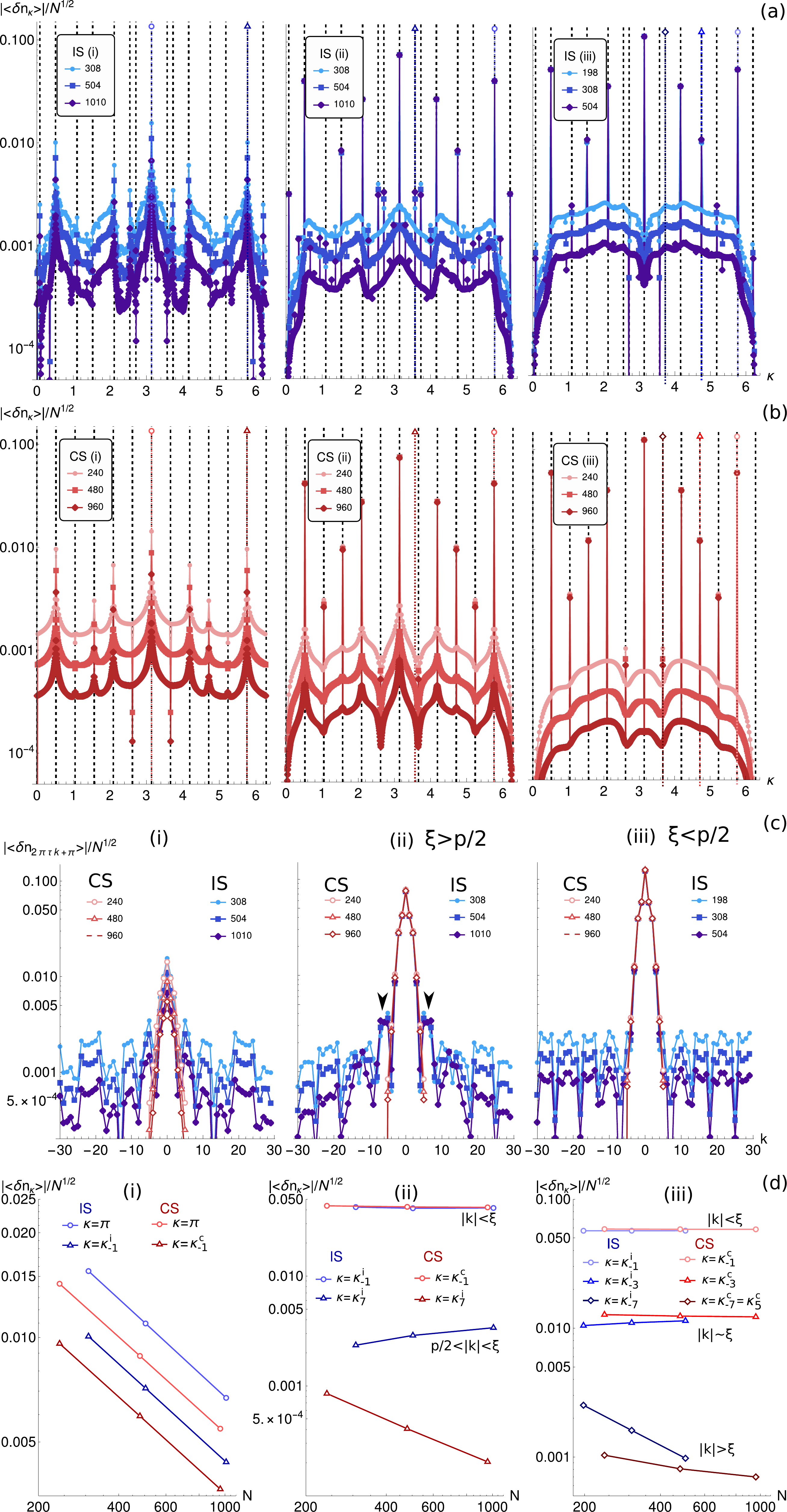}\caption{Fourier transform of charge fluctuations, $|\langle\delta n_{\kappa}\rangle|/N^{1/2}$,
for IS (a) and CS (b), for $V_{2}=0.975$ and for the values of $U$
marked in Fig.$\,$\ref{fig:LL-CDW}(d). The dashed lines correspond
to $\kappa_{k}^{i}=\pi+2\pi\tau k,k=-7,\cdots,7$ in (a) and to $\kappa_{k}^{c}=\pi+2\pi\tau_{c}k,k=-6,\cdots,5$
in (b) - in the latter case, the wave vectors repeat for $k$ outside
this interval. (c) $|\langle\delta n_{\kappa}\rangle|/N^{1/2}$ plotted
as a function of the ordering index $k$. For $\xi>p/2$, the IS has
order at wave vectors that are absent in the CS, marked by the arrows
in the middle figure. For $\xi<p/2$ on the other hand there is only
order at $|k|\lesssim\xi<p/2$ both in the CS and IS. (d) Finite-size
scaling for selected peaks, indicated in the legend and by the open
markers in (a,b). For $U<U_{c}(V_{2})$ (i), $|\langle\delta n_{\kappa}\rangle|/N^{1/2}\rightarrow0$,
which implies that $K_{\kappa}=0$ for all $\kappa$. For $U>U_{c}(V_{2})$,
$K_{\kappa}>0$ for $|k|\lesssim\xi/2$. For $U\approx U_{c}(V_{2})$
(ii), we can have $\xi>p/2$ in which case $K_{\kappa}$ becomes finite
for some $|k|>p/2.$ An example is shown for $k=7$, where we see
that $K_{\kappa}>0$ for $\kappa=\kappa_{7}^{i}\protect\neq\kappa_{-5}^{i}$
for the IS, while no equivalent wave vector exists for the CS since
$\kappa_{7}^{c}=\kappa_{-5}^{c}$. In fact we see that $K_{\kappa}\approx0$
for $\kappa=\kappa_{7}^{i}$ in the CS. For larger $U$ (iii), we
see that $K_{\kappa}>0$ for $|k|\lesssim\xi<p/2$, while $K_{\kappa}\approx0$
for $|k|>\xi$. \label{fig:LL-CDW_dnk}}
\end{figure}

 To compute the phase transition points in Fig.$\,$\ref{fig:main}(c),
for the incommensurate case, we used a thermodynamic-limit extrapolation
of the finite-size results for each computed quantity and estimated
the critical point to be the average of the obtained values (see SM
\citep{SM} for details). For the commensurate case, a thermodynamic-limit
extrapolation of the ${\rm IPR}_{\kappa}(\langle\delta\bm{n}\rangle)$
results was used to obtain the phase diagram in the inset of Fig.$\,$\ref{fig:main}(c).

\paragraph*{Fate of extended states in the presence of interactions. ---\label{sec:LLCDW}}

We start by establishing the stability of the (ballistic and gapless)
LL state, away from $U=0$, when interactions are turned on and $V_{2}<1$.
This is shown in Figs.$\,$\ref{fig:LL-CDW}(a) and (c) where, for
$U<U_{c}(V_{2})$, both $\Delta_{c}$ and ${\rm IPR}_{\kappa}(\langle\delta\bm{n}\rangle)$
decrease linearly with $N$. Interestingly, Fig.$\,$\ref{fig:LL-CDW}(a)
shows that $\lim_{N\to\infty}N\Delta_{c}$ decreases rapidly as $V_{2}\to1$,
which we interpret as a many-body signature of the flattening of the
dispersion relation observed at in Fig.$\,$\ref{fig:main}(a) for
$U=0$.

For $U>U_{c}(V_{2})$, the results of Figs.$\,$\ref{fig:LL-CDW}(a)
and (c) are compatible with $\lim_{N\to\infty}\Delta_{c}\neq0$ and
$\lim_{N\to\infty}{\rm IPR}_{\kappa}(\langle\delta\bm{n}\rangle)\neq0$,
i.e. with the existence of a gapped state with CDW order. The LL-CDW
phase transition at $U_{c}(V_{2})$ is well captured by the divergence
of $\chi_{F}$ with $N$, shown in Fig.$\,$\ref{fig:LL-CDW}(b).

Fig.$\,$\ref{fig:LL-CDW}(c)-right panel shows that the stability
of the LL phase holds for both CS and IS and that the LL-CDW transition
arises for similar $U_{c}(V_{2})$. However, within the CDW we observe
a subtle difference between the CS and IS sufficiently close to $U_{c}(V_{2})$.

Fig.$\,$\ref{fig:LL-CDW_dnk}(c) depicts the normalized Fourier transform
of the charge fluctuations $|\langle\delta n_{\kappa}\rangle|$ for
the IS (a) and CS (b), for $V_{2}\lesssim1$. Fig.$\,$\ref{fig:LL-CDW_dnk}(c)
shows the scaling of $|\langle\delta n_{\kappa}\rangle|/N^{1/2}$
for the some chosen values of the momentum $\kappa$.

In the disordered LL phase {[}case (i){]}, $K_{\kappa}=\lim_{N\to\infty}|\langle\delta n_{\kappa}\rangle|/N^{1/2}$
vanishes for all $\kappa$, whereas for the staggered CDW at $V_{2}=0$
(not shown) $K_{\kappa=\pi}=1$ and $K_{\kappa\neq\pi}=0$. For $V_{2}>0$,
in the ordered state there are contributions from several $\kappa$.

For CS with period $p$ there are $p$ inequivalent contributing wave
vectors. For the example of Fig.$\,$\ref{fig:LL-CDW_dnk}(b), with
$\tau=\tau_{c}=7/12$, there are $p=12$ wave vectors, $\kappa_{k}^{c}=\pi+2\pi\tau_{c}k$
with $k=-6,...,5$ (shown as dashed lines).

For IS, the number of contributing wave vectors increases linearly
with $N$, and we find they are given by $\kappa_{k}^{i}=\pi+2\pi\tau k,\textrm{ }k=0,\pm1,\pm2,...,\pm N/2$.

In Fig.$\,$\ref{fig:LL-CDW_dnk}(c), the contributions of each wave-vector
are ordered by magnitude, $K_{\kappa_{\abs k}}>K_{\kappa_{\abs k+1}}$.
For both IS and CS, $K_{\kappa_{\abs k}}$ decreases exponentially
with $k$ . i.e. $K_{\kappa_{\abs k}}\sim e^{-\abs k/\xi}$, where
$\xi$ is a characteristic decay-scale that decreases with increasing
$U>U_{c}(V_{2})$. Therefore, we may distinguish two cases: $\xi>p/2$
and $\xi<p/2$. For $\xi\ll p/2$, {[}case (iii) of figure \ref{fig:LL-CDW_dnk}(c){]}
there is no difference between the IS and CS since all relevant contributing
wave-vectors have $\abs k\lesssim\xi\ll p/2$. On the other hand,
for $U$ sufficiently close to $U_{c}(V_{2})$ and $V_{2}$ sufficiently
close to $1$, we can find a region where $\xi>p/2$ {[}as illustrated
in case (ii) of figure \ref{fig:LL-CDW_dnk}(c){]}. For this case there
are more relevant contributing wave-vectors in IS than for CS and
therefore the ${\rm IPR}_{\kappa}(\langle\delta\bm{n}\rangle)$ differs
for these two cases. Nevertheless the exponential decay of $K_{\kappa_{\abs k}}$
ensures ${\rm IPR}_{\kappa}(\langle\delta\bm{n}\rangle)$ is finite
and qualitatively similar for CS and IS.

As discussed in the methods section, results of Figs.\ref{fig:LL-CDW,fig:LL-CDW_dnk}
are obtained with DMRG by applying a symmetry-breaking field. In the
SM \citep{SM}, we analyzed the structure factor with no symmetry-breaking
while imposing closed boundary conditions. While the available system
sizes are significantly smaller, the results seem confirm the findings
in the main text (see SM \citep{SM}).

\paragraph*{Fate of critical multifractal states in the presence of interactions.
---\label{sec:CritCDW}}

\begin{figure}[H]
\centering{}\includegraphics[width=1\columnwidth]{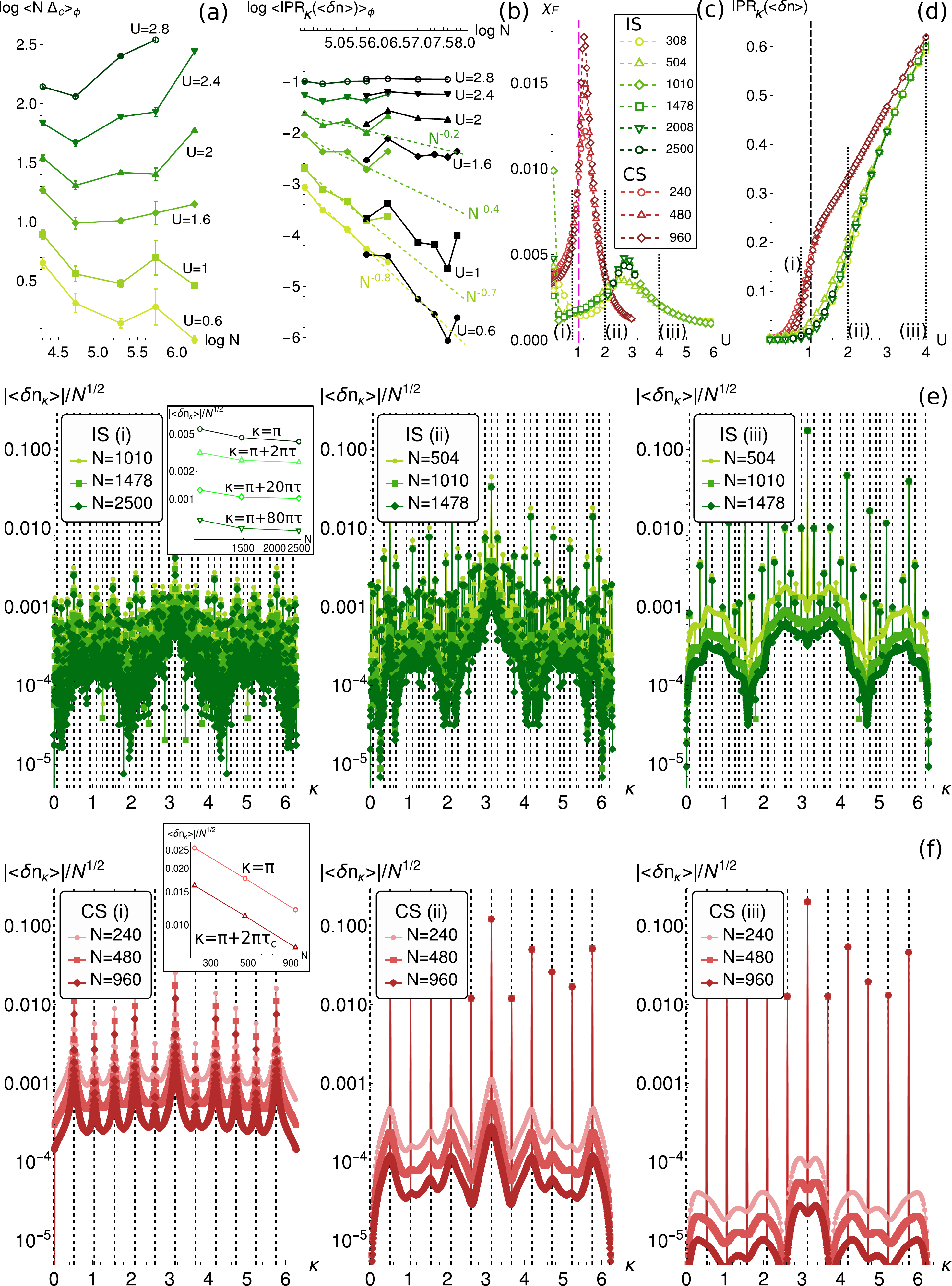}\caption{(a,b) Results after averaging over different configurations of $\phi$,
for sizes up to $N=504$, for the IS. Most results were obtained for
$N_{c}=50$ configurations of $\phi$, given by $\phi_{i}=2\pi i/N_{c},\textrm{ }i=0,\cdots,N_{c}-1$.
In black we show results obtained for larger sizes, for fixed $\phi=1.8$.
Dashed lines correspond to linear fits using only the 4 smallest system
sizes. (c,d) Results for IS: green, for fixed $\phi=1.8$ and larger
sizes (up to $N=2500$); CS: red, for fixed $\phi=0$. The dashed
line corresponds to the extrapolated critical point for the CS using
the ${\rm IPR}_{k}(\langle\delta\bm{n}\rangle)$ results. (e,f) Examples
of $|\langle\delta n_{\kappa}\rangle|/N^{1/2}$ for the $U$ values
marked in (c,d) for the IS (e) and the CS (f). The dashed lines correspond
to $\kappa=\pi+2\pi\tau k,k\in[-16,16]$ in (e) and to $\kappa=\pi+2\pi\tau_{c}k,k\in]-6,6]$
in (f).\label{fig:crit-CDW}}
\end{figure}

We now turn to the region $V_{2}>1$ that hosts critical multifractal
states for $U=0$. Here, upon increasing $U,$we find that incommensurability
plays a crucial role as CS and IS give rise to radically different
physics. This is illustrated in Figs.$\,$\ref{fig:crit-CDW}(c) and
(d) that show the behavior of $\chi_{F}$ and ${\rm IPR}_{\kappa}(\langle\delta\bm{n}\rangle)$.
For CS, the finite size scaling of $\chi_{F}$ around $U=U_{c}\approx1$
indicates a thermodynamic divergence concomitant with an abrupt change
of ${\rm IPR}_{\kappa}(\langle\delta\bm{n}\rangle)$. Together with
the behavior for the entanglement entropy (see SM \citep{SM}), the
overall physical picture is similar to that obtained for $V_{2}<1$--
the LL is stable at small $U$ and transitions to a CDW for a critical
value of the interactions $U=U_{c}\left(V_{2}\right)$. Note that
for CS with fixed $\phi$ and $V_{2}\geq1$ there are cases for which
hoppings vanish within a unit cell signaling that the narrow-band
becomes exactly flat at $U=0$ (indicated by the dot-dashed lines
in the inset of Fig.$\,$\ref{fig:main}(c), for $\phi=0$). Although
the $U_{c}$ of the LL-CDW transition is suppressed around these points,
it remains finite.

On the other hand, for IS, both $\chi_{F}$ and ${\rm IPR}_{\kappa}(\langle\delta\bm{n}\rangle)$
converge everywhere (for the larger system sizes) which suggest no
phase transition arises for $U>0$. Instead, our results point to
a crossover, signaled by a peak of $\chi_{F}$ around $U\sim2.7$,
from a regular CDW state at large $U$ to a peculiar CDW -- the quasi-fractal
CDW -- at small $U$. These results are further corroborated by those
in Figs.$\,$\ref{fig:crit-CDW}(a) and (b). Fig.$\,$\ref{fig:crit-CDW}(a)
shows that for larger values of $U$ the average $N\Delta_{c}$ increases
with $N$(for sufficiently large $N$) but remains essentially constant
for $U\lesssim1$. In the same way, ${\rm IPR}_{\kappa}(\langle\delta\bm{n}\rangle)$
decreases with $N$ for $U\lesssim1$, as see Fig.$\,$\ref{fig:crit-CDW}(b).
Note that for $V_{2}\gtrsim1$ there is a dependence on the phase
$\phi$ for the smaller system sizes. For those we average over different
configurations of $\phi$, while for larger $N$, when this dependence
is small, we only show a typical value of $\phi$. The results in
Figs.$\,$\ref{fig:crit-CDW}(a-d) point to an instability of critical
multifractal states, towards a CDW phase with a very small gap and
a large number of wave vectors contributing to the charge fluctuations.
The quasi-fractal nature of this regime can be seen from the scaling
${\rm IPR}_{\kappa}(\langle\delta\bm{n}\rangle)\sim N^{-\alpha}$,
$0<\alpha<1$ {[}see fits in Fig.$\,$\ref{fig:crit-CDW}(b){]}.

It is worth noting that in the vicinity of $V_{2}=1$, the flattening
of the narrow-band suppresses the charge gap, and makes it impossible
to meet the DMRG convergence criteria. This is particularly problematic
at small $U$, where $\Delta_{c}$ takes the smallest values. To circumvent
this problem we consider $V_{2}=3.5$ throughout this section, for
which we were able to reach DMRG convergence criteria down to $U=0$
(see SM \citep{SM} ). Nevertheless, the conclusions above are consistent
with the results obtained for smaller values of $V_{2}$ (down to
$V_{2}=1$) given in the SM \citep{SM} for which convergence can
be only be achieved for sufficiently large $U$. Furthermore, the
vanishing of the critical interaction strength $U_{c}$ when the critical
regime is reached is also supported by the observed behavior $U_{c}\left(V_{2}\right)\sim(1-V_{2})^{\mu}$,
with $\mu\approx0.4$ for $V_{2}<1$ (see SM \citep{SM}).

The characteristic interaction strength where the quasi-fractal CDW
crosses over to regular CDW order cannot be defined uniquely. In Fig.$\,$\ref{fig:main}
(dashed lines), we use the points at which $\pd{\rm IPR}_{\kappa}(\langle\delta\bm{n}\rangle)/\pd U$
is maximal. An alternative definition, consisting of maximizing $\chi_{F}$,
yields slightly larger values of $U$ but the same physical picture.

To further illustrate how quasi-fractal CDW order differs from previously
known states, we show in Figs.$\,$\ref{fig:crit-CDW}(e,f) representative
results for $|\langle\delta n_{\kappa}\rangle|/N^{1/2}$ in IS and
CS. For large enough $U$, there are no significant differences between
the type of order observed in IS {[}see (e-iii){]}, and CS {[}see
(f-iii){]}. In this case $K_{\kappa_{\abs k}}\sim e^{-\abs k/\xi}$
, as for $V_{2}<1$ (further details on the decay are given in the
SM \citep{SM}).

This picture drastically changes for the IS when $U$ crosses-over
to the quasi-fractal regime. Here, the IS order acquires contributions
from a large number of wave-vectors as shown in (e-ii) and in the
SM \citep{SM}. As $U$ decreases within the quasi-fractal regime,
$\xi$ increases and rapidly becomes larger than the system size,
$\xi\gtrsim N$. This is illustrated in (e-i) where a finite contribution
$K_{\kappa_{k}}$ is shown to persist up to large $|k|$. Since ${\rm IPR}_{\kappa}(\langle\delta\bm{n}\rangle)\rightarrow0$
as $U\rightarrow0$, we conjecture that $\xi\to\infty$ in the limit
$U\to0$, with all wave-vectors, $\kappa=\kappa_{k}^{i}$, contributing
for $U=0^{+}$.

This is in sharp contrast with CS for which contributions only arise
from the $p$ wave-vectors $\kappa=\kappa_{k}^{c}$ {[}see (f-ii){]}
within the ordered phase. For sufficiently small $U$, CS enter the
disordered phase signaled by the decay of $|\langle\delta n_{\kappa}\rangle|/N^{1/2}$
with $N$ in (f-i), which implies $K_{\kappa}=0$ for all $\kappa$.

\paragraph{Discussion.---}

We determined the ground-state phase diagram of a 1D interacting narrow-band
moiré system, at half-filling, which hosts both extended and critical
single-particle states. Once interactions are switched on, commensurate
(periodic) systems always stabilize the (gapless) Luttinger liquid
phase for sufficiently small values of interaction strength, with
a transition to a (gapped) periodic moiré CDW state occurring at a
finite $U$ value.

Contrastingly, in the incommensurate (quasiperiodic) system the Luttinger
liquid phase is suppressed immediately for an infinitesimal interaction
strength when the non-interacting states are critical and multifractal.
The ensuing ordered phase is a quasi-fractal CDW, characterized by
contributions from a large number of wave vectors. Increasing the
interaction strength induces a crossover from the quasi-fractal CDW
regime, to a regular quasiperiodic moiré regime, in which only a small
subset of wave vectors effectively contribute, as in the commensurate
case.

These findings demonstrate that quasiperiodicity may radically change
the ground state properties of 1D moiré systems with respect to their
periodic counterparts. Perhaps our most important result is that,
while the narrow band plays an important role in enhancing interactions
both for commensurate and incommensurate structures, only in the quasiperiodic
case the ordered phase extends down to infinitesimal interactions,
resulting in infinitely many contributing wave vectors. This quasi-factal
regime cannot be stabilized in the commensurate case even in the presence
of a narrow band.

Our results are a significant first step in understanding the interplay
between incommensurability and interactions in moiré systems, clearly
demonstrating the need to explicitly consider quasiperiodicity. They
also suggest that other critical multifractal phases of non-interacting
systems can yield exotic correlated states once interactions are turned
on. A particular promising avenue is to study the role of interactions
in the recently theoretically reported critical phases of tBLG \citep{Goncalves_2022_2DMat,Fu2020,PhysRevB.101.235121}
and in other 2D moiré systems such as twisted trilayer graphene where
strong quasiperiodic effects were recently experimentally reported
\citep{uri2023superconductivity}, to understand if incommensurability
effects can help explain the plethora of exotic phases shown to arise
in these systems.

\label{sec:Discussion}
\begin{acknowledgments}
The authors MG and PR acknowledge partial support from Fundação para
a Ciência e Tecnologia (FCT-Portugal) through Grant No. UID/CTM/04540/2019.
BA and EVC acknowledge partial support from FCT-Portugal through Grant
No. UIDB/04650/2020. MG acknowledges further support from FCT-Portugal
through the Grant SFRH/BD/145152/2019. BA acknowledges further support
from FCT-Portugal through Grant No. CEECIND/02936/2017. Some computations
were performed on Tianhe-2JK cluster at the Beijing Computational
Science Research Center (CSRC)
\end{acknowledgments}

\bibliographystyle{apsrev4-1}
\bibliography{Quasiperiodic_Interactions,1D_Hidden_SD_Paper}

%%%%%%%%%%%%%%%%%%%%%%%%%%%%%%%%%%%%%%%%%%%%%%%%%%%%%%%%%

%%%%%%%%%%%%%%%%%%%%%%%%%%%%%%%%%%%%%%%%%%%%%%%%%%%%%%%%%

%%%%%%%%%%%%%%%%%%%%%%%%%%%%%%%%%%%%%%%%%%%%%%%%%%%%%%%%%

\clearpage\onecolumngrid

\beginsupplement
\begin{center}
\textbf{\large{}Supplemental Material for: \vspace{0.1cm}
}{\large\par}
\par\end{center}

\begin{center}
{\Large{}Incommensurability-driven quasi-fractal order in 1D narrow-band
moiré system}{\Large\par}
\par\end{center}

\vspace{0.3cm}

\tableofcontents{}

\section{Origin of narrow-band}

To see how a narrow-band develops around $E=0$ for the model in Eq.$\,$\ref{eq:Hamiltonian}
in the non-interacting limit, we analyse the quasiperiodic hopping
term in momentum space, which reads
\begin{equation}
H_{0}^{\text{qp}}=V_{2}\sum_{n}\cos[2\pi\tau(n+1/2)+\phi]c_{n}^{\dagger}c_{n+1}+\textrm{h.c.}
\end{equation}
Making a Fourier transform in the electron operators,

\begin{equation}
c_{n}=\frac{1}{\sqrt{N}}\sum_{k}e^{ikn}c_{k},
\end{equation}
we have that

\begin{equation}
\begin{aligned}H_{0}^{\text{qp}}=\sum_{k}\cos(k+\pi\tau)[e^{i\phi}c_{k+2\pi\tau}^{\dagger}c_{k}+e^{-i\phi}c_{k}^{\dagger}c_{k+2\pi\tau}]\end{aligned}
.\label{eq:flatband_origin}
\end{equation}

\begin{figure}[H]
\begin{centering}
\includegraphics[width=0.4\columnwidth]{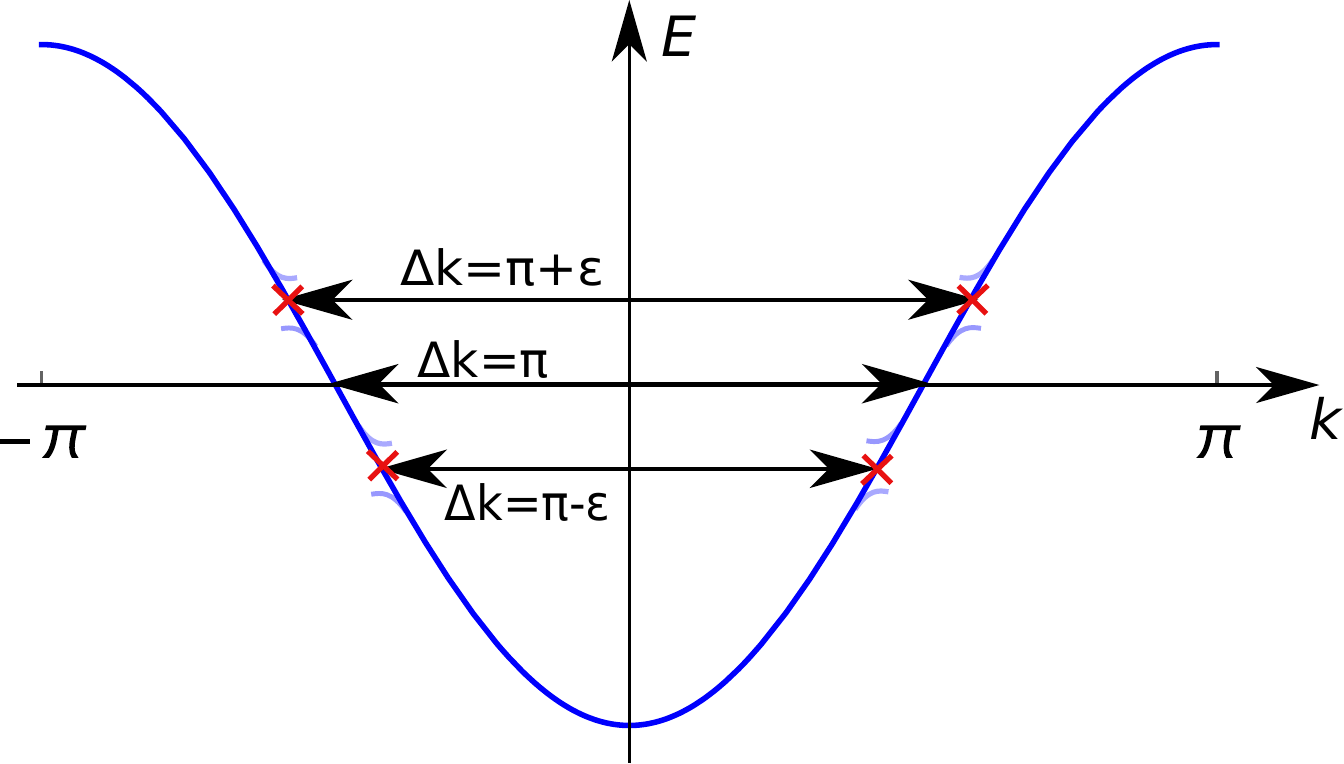}
\par\end{centering}
\caption{Narrow-band formation: momenta separated by $\Delta k=\pi\pm\epsilon$
are coupled and gaps are opened, leaving a narrow-band around $E=0$.
\label{fig:Flatband_formation}}
\end{figure}

From this calculation we see that the quasiperiodic hopping term
couples Bloch momenta $k$ and $k\pm2\pi\tau n$ in $n$-th order
perturbation theory. For small $V_{2}$ we can stick to first order
and, setting $\tau=1/2+\epsilon$, momenta with $\Delta k=\pi\pm\epsilon$
will be coupled. As a consequence, gaps open at momenta separated
by $\Delta k$, leaving a narrow-band around $E=0$ that is flatter
for smaller $\epsilon$ (and at the same time corresponds to a smaller
portion of the spectrum), see Fig.$\,$\ref{fig:Flatband_formation}.

\section{Finite-size scalling analysis}

In this section we provide details on how the critical points in the
phase diagram in Fig.$\,$\ref{fig:main} were obtained. Using finite-size
scalling analysis, we computed thermodynamic-limit extrapolations
for several different quantities. Besides the charge gap, fidelity
susceptibility and ${\rm IPR}_{\kappa}(\langle\delta\bm{n}\rangle)$
shown in the main text, we also computed the entanglement entropy
at the middle bond, $S$, and the quantity $\chi_{R}=N^{-1}\sum_{i}\langle\delta n_{i}\rangle^{2}$.

Below, we detail how the thermodynamic limit extrapolations were carried
out for each quantity.

\begin{figure}[h]
\begin{centering}
\includegraphics[width=0.6\columnwidth]{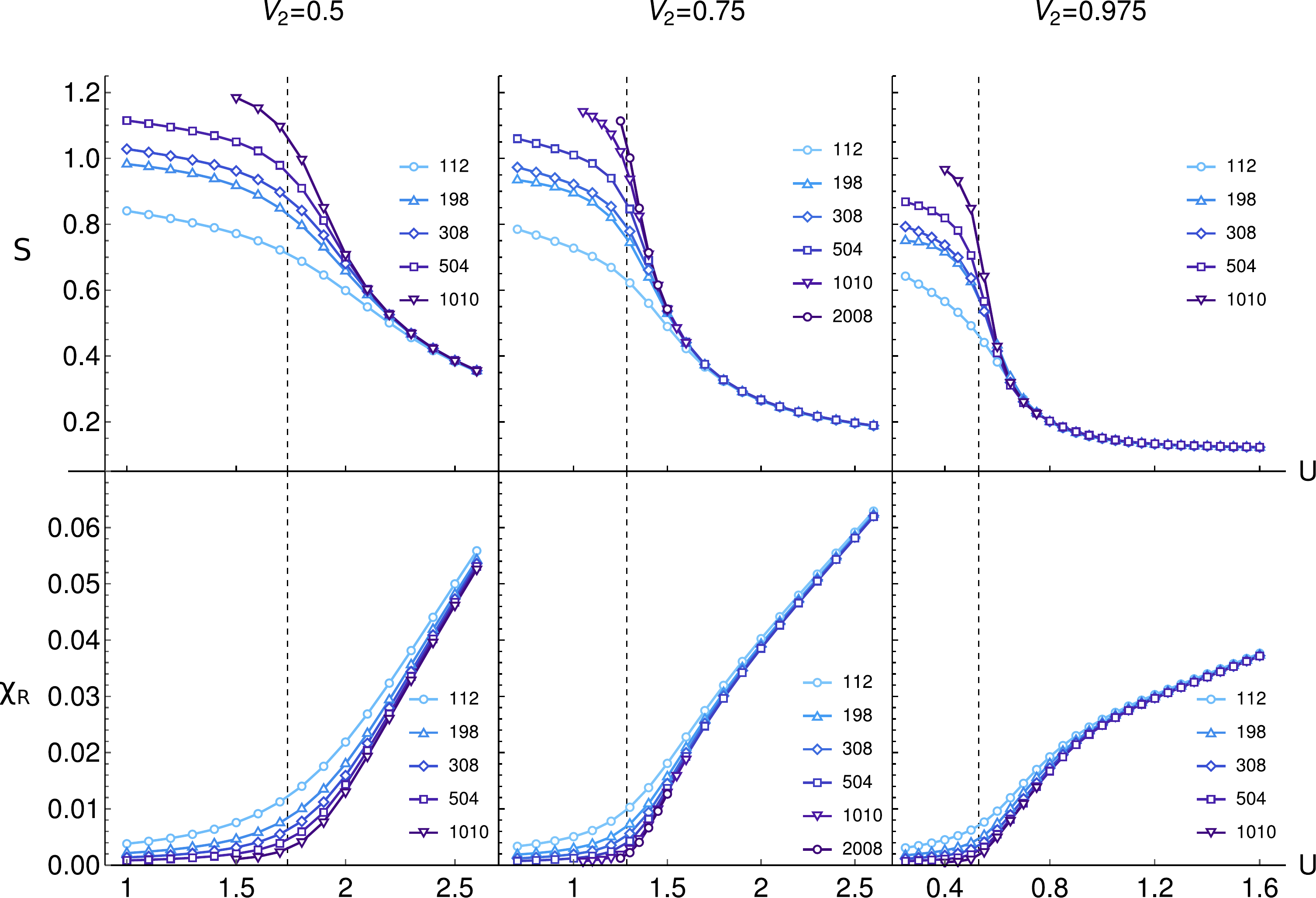}
\par\end{centering}
\caption{Finite-size scaling results for (a) entanglement entropy at middle
bond, $S$; (b) $\chi_{R}=N^{-1}\sum_{i}\langle\delta n_{i}\rangle^{2}$.
The top label indicates the value of $V_{2}$ used in the corresponding
column, in (a,b). The dashed lines in each figure correspond to the
thermodynamic limit extrapolation of the critical point, that we detail
below. \label{fig:S_chiR}}
\end{figure}

\paragraph{Charge gap.---}

\begin{figure}[H]
\centering{}\includegraphics[width=0.6\columnwidth]{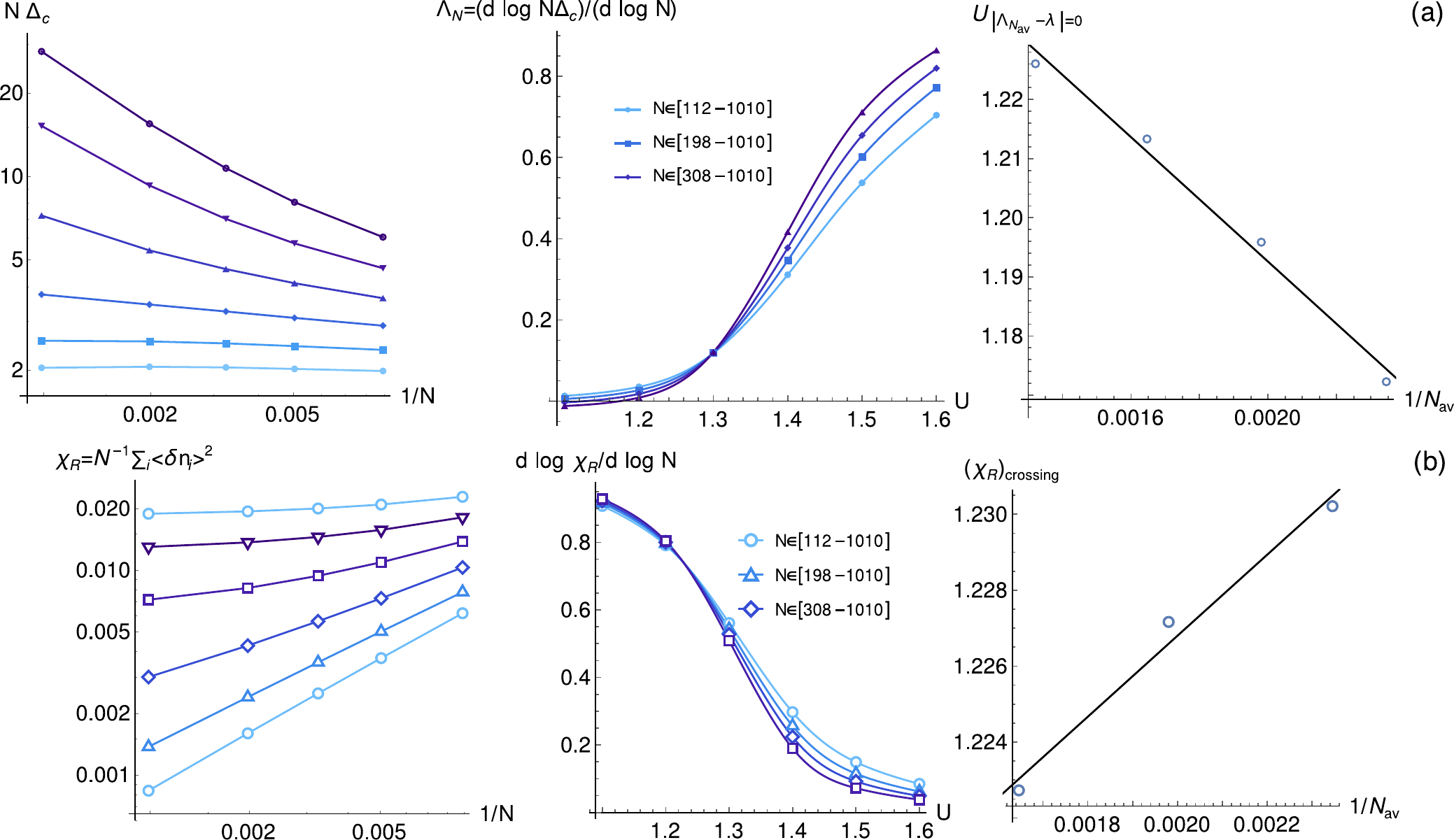}\caption{Example of thermodynamic-limit extrapolations for the charge gap and
$\chi_{R}$, for $V_{2}=0.75$. \label{fig:Cgap_and_chiR}}
\end{figure}

For large enough system and close to the phase transition,we assume
that the charge gap obeys the scalling hypothesis
\begin{equation}
N\Delta_{c}(U,N)=N^{a}f\left(N^{b}\left[U-U_{c}\right]\right),
\end{equation}
where $f(x)$ is a scalling function, $a$ and $b$ are some scalling
exponents, and $U_{c}$ is the critical interaction strength. Taking
a logarithmic derivative of $N\Delta_{c}(U,N)$ with the system size
we have that
\begin{equation}
\frac{d\log\left(N\Delta_{c}(U,N)\right)}{d\text{log}N}=a+\frac{f^{\prime}\left(N^{b}\left[U-U_{c}\right]\right)}{f\left(N^{b}\left[U-U_{c}\right]\right)}N^{b}\left[U-U_{c}\right].
\end{equation}
From the above expression we see that at $U=U_{c}$, the above quantity
becomes independent of the system size $N$, and by ploting it versus
$U$ the lines for different system sizes should cross. However, we
have not used this criterion for the charge gap, for two reasons:
(i) it provides a poor result for the $V_{2}=0$ case, for which the
$U_{c}$ is know, (ii) for some values of $V_{2}$, there is no clear
crossing point. So instead, we notice that by solving the condition
\begin{equation}
\frac{d\log\left(N\Delta_{c}(U,N)\right)}{d\text{log}N}=\lambda
\end{equation}
as a function of $U$, for some fixed value $\lambda$ close to 0,
under the scalling hypothesis, this is equivalent to solving
\begin{equation}
a+\frac{f^{\prime}\left(x_{\lambda}^{*}\right)}{f\left(x_{\lambda}^{*}\right)}x_{\lambda}^{*}=\lambda,
\end{equation}
where $x_{\lambda}^{*}=N^{b}\left[U_{\lambda}^{\text{*}}(N)-U_{c}\right]$
and $U_{\lambda}^{\text{*}}(N)$ is the value of $U$ such that the
logarithimic derivative equals $\lambda$ for a given system size.
Therefore, under the scalling hypothesis we have 
\[
U_{\lambda}^{\text{*}}(N)=U_{c}+\frac{x_{\lambda}^{*}}{N^{b}}
\]
and by determining $U_{\lambda}^{\text{*}}$ for different system
sizes and extrapolating it to $N\rightarrow\infty$, we can extract
$U_{c}$. We numerically implemented this criterion as follows. For
each $U$ we make a linear fit of the $\log N\Delta_{c}$ data points
as a function of $\log N$, excluding the smallest sizes up to when
only two sizes remain. The slopes obtained by fitting each set of
system sizes give rise to the data points $(U,\Lambda_{N_{{\rm av}}})$,
where $\Lambda_{N_{{\rm av}}}=\frac{{\rm d}\log N\Delta_{c}}{{\rm d}\log N}\big|_{N=N_{{\rm av}}}$
and $N_{{\rm av}}$ is the average of the system sizes used to make
the fit. These results are exemplified in Fig.$\,$\ref{fig:Cgap_and_chiR}(a-middle).
In the next step we interpolate these points and check when the resulting
curves cross a value $\lambda$ that we choose close to $0$ (in Fig.$\,$\ref{fig:Cgap_and_chiR}
we used $\lambda=0.025$). The reason for this choice is that choosing
$\lambda=0$ gives a very accurate estimation of the well-known critical
point $U_{c}=2$ for $V_{2}=0$ ($U_{c}=2.001\pm0.008$). We then
extrapolate the value of $U$ for which $\Lambda_{N_{{\rm av}}}=\lambda$,
that we label $U_{\lambda}^{\text{*}}(N)$, and plotted it as a function
of $N^{-1}$, observed that the exponent $b\approx1$ and extracted
$U_{c}$ by making a linear fit of $U_{\lambda}^{\text{*}}(N)$ vs
$N^{-1}$.We notice that in this method, we cannot always choose $\lambda=0$,
since in some cases the $\Lambda_{N_{{\rm av}}}$ curves do not cross
this value.

\paragraph{$\chi_{R}$.---}

For the quantity $\chi_{R}$, we expect it to scale to zero with $N$
in the LL phase (the fluctuations $\langle\delta n_{i}\rangle$ are
non-extensive) and to converge to a constant in the CDW phase (due
to extensive fluctuations). An example of this hevaviour is shown
in \ref{fig:S_chiR}(b). We made the same scalling hypothesis for
$\chi_{R}$ as for the charge gap and used an identical procedure
to determine the logarithmic derivative of $\chi_{R}$ with respect
to system size, $\Lambda_{N_{{\rm av}}}^{R}=\frac{{\rm d}\log\chi_{R}}{{\rm d}\log N}\big|_{N=N_{{\rm av}}}$.
To extrapolate to the thermodynamic limit, we used the criterion of
the $\Lambda_{N_{{\rm av}}}^{R}$ at the phase transition, by extracting
crossing of the curves of $\Lambda_{N_{{\rm av}}}^{R}$ vs $U$ for
different system sizes, as this was always well defined. An example
is shown in Fig.$\,$\ref{fig:Cgap_and_chiR}(b).

\paragraph{Fidelity susceptibility.---}

For $\chi_{F}$ we extrapolate the maximizant $U_{c}(N)$ using the
3 largest sizes and assuming $U_{c}(N)=U_{c}+\beta/N$.

\paragraph{Entanglement entropy.---}

In the case of the entanglement entropy, it increases with $N$ in
the LL phase, while it converges to a constant in the CDW phase, as
illustrated in Fig.$\,$\ref{fig:Cgap_and_chiR}(a). For finite $N$,
$|\pd S/\pd U|$ has a maximum at a value $U^{*}(L)$ that approaches
the critical interaction strength $U_{c}$ as $N$ is increased. We
extrapolated the maximizant of $|dS/dU(N)|$ using the 3 largest sizes,
again assuming that $U^{*}(L)=U_{c}+\beta/N$..

\paragraph{${\rm IPR}_{\kappa}(\langle\delta\bm{n}\rangle)$.---}

\begin{figure}[H]
\centering{}\includegraphics[width=0.6\columnwidth]{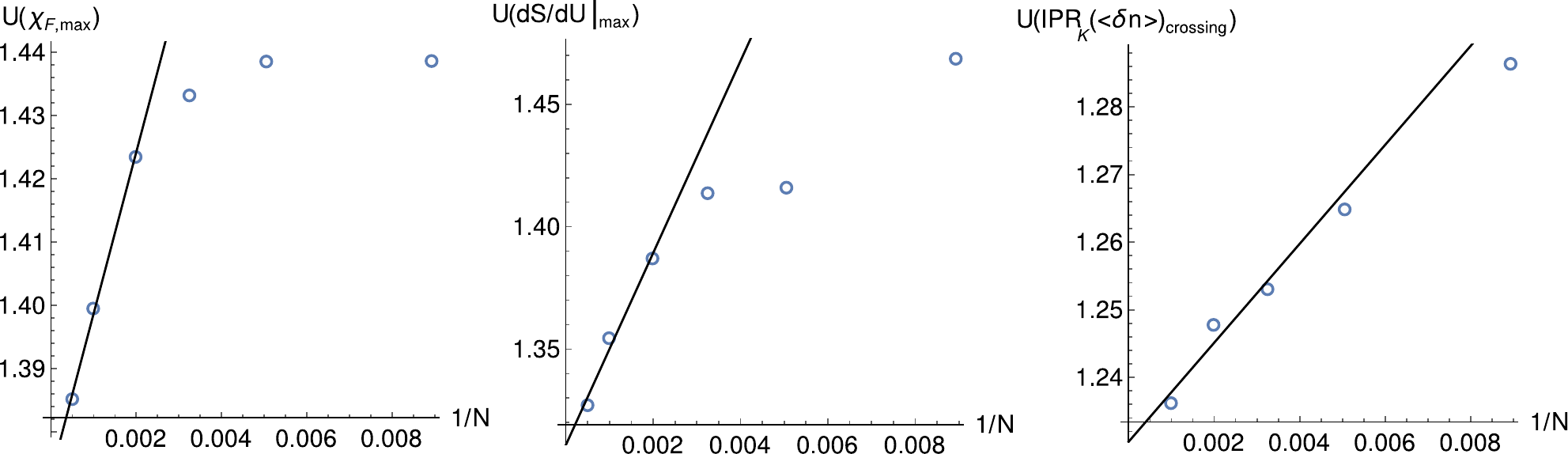}\caption{Example of thermodynamic-limit extrapolations for $\chi_{F}$, $dS/dU|_{{\rm max}}$
and ${\rm IPR}_{\kappa}(\langle\delta\bm{n}\rangle)$, for $V_{2}=0.75$.
 \label{fig:remain_extraps}}
\end{figure}

For ${\rm IPR}_{\kappa}(\langle\delta\bm{n}\rangle)$ we extrapolate
the crossing point between the curves of consecutive system sizes,
considering the crossings for the 5 largest system sizes. Extrapolations
for these quantities are exemplified in Fig.$\,$\ref{fig:remain_extraps},
for $V_{2}=0.75$.

\begin{figure}[H]
\begin{centering}
\includegraphics[width=0.25\columnwidth]{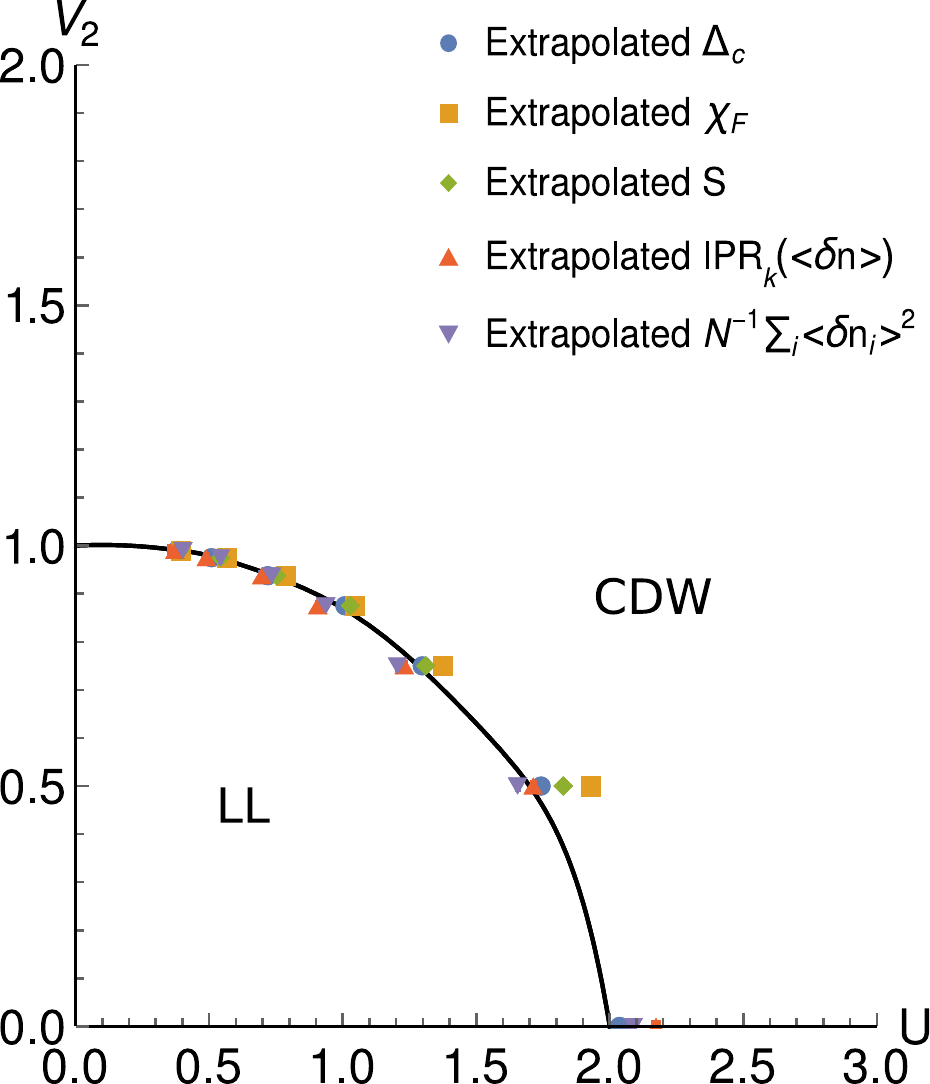}
\par\end{centering}
\caption{Extrapolated critical points for the different quantities. \label{fig:sup_extrap}}
\end{figure}

Finally, in Fig.$\,$\ref{fig:sup_extrap}, we show the calculations
of the extrapolated critical points, for each different quantity,
separately. We can see that for lower $V_{2}$, the extrapolation
using $\chi_{F}$ considerably overestimates the critical point comparing
to the remaining quantities. This suggests that the finite-size effects
are more severe for this quantity, in this regime. For $V_{2}=0.5$,
we therefore estimate the critical point by averaging over all results,
except for the $\chi_{F}$ extrapolation. On the other hand, for $V_{2}\geq0.75$
we average over all results to estimate the critical point.

\begin{figure}[H]
\centering{}\includegraphics[width=0.5\columnwidth]{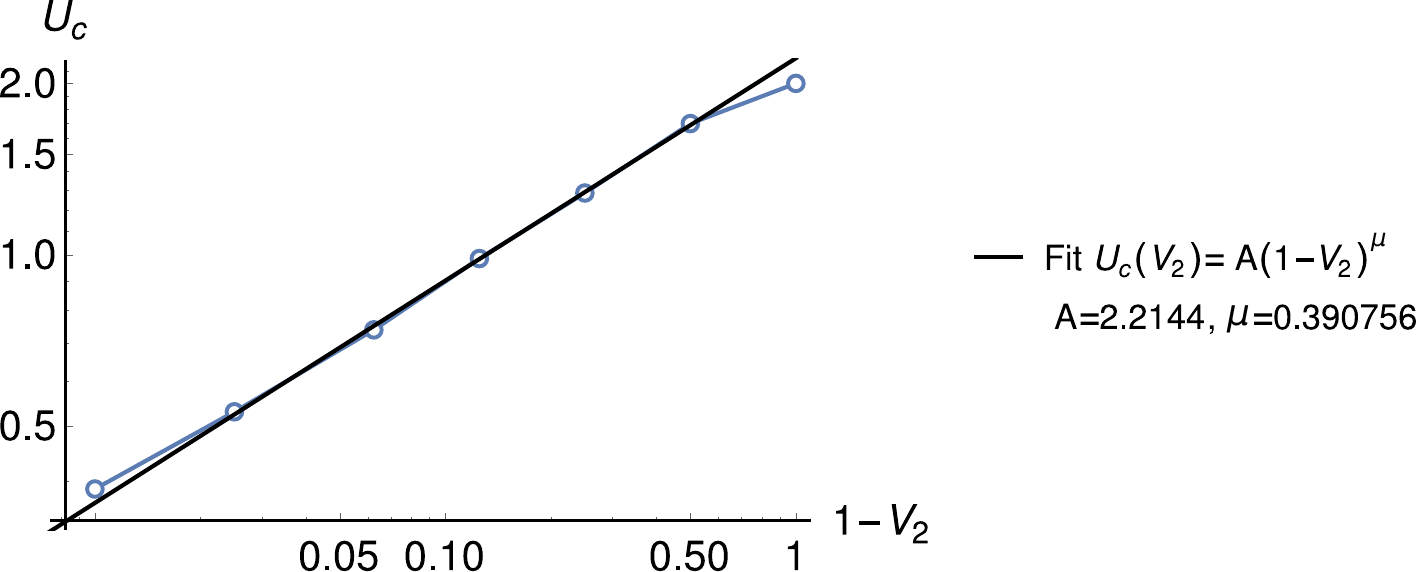}\caption{Fit of $\log U_{c}$ vs. $\log(1-V_{2})$, using the 5 points with
larger $V_{2}$.\label{fig:Uc_vs_1-V2}}
\end{figure}

We finish this section by supporting the claim made in the main text
that $U_{c}\sim(1-V_{2})^{\mu}$, with $\mu\approx0.4$. The results
are presented in Fig.$\,$\ref{fig:Uc_vs_1-V2}.

\section{Additional results for $V_{2}\protect\geq1$}

\begin{figure}[h]
\begin{centering}
\includegraphics[width=0.85\columnwidth]{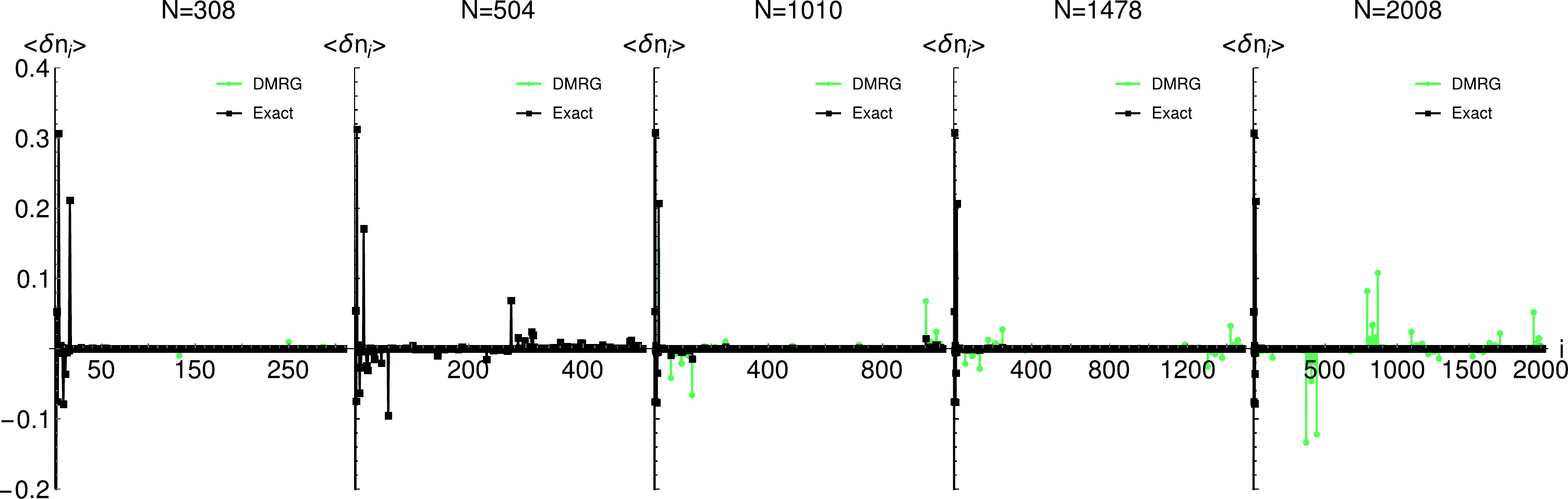}
\par\end{centering}
\caption{Comparison of exact and DMRG calculations for $\langle\delta n_{i}\rangle$,
for $U=0,V_{2}=3.5,\phi=1.8$ and different system sizes.\label{fig:U-0_V2-3.5}}
\end{figure}

In this section we present additional results inside the region $V_{2}\geq1$
of the phase diagram. In the main text we chose $V_{2}=3.5$, so that
the convergence of the DMRG calculations was satisfactory down to
$U=0$ for the used system sizes. As we have seen there, for a given
system size, the smallest charge gaps occur for smaller $U$, in which
case a good enough convergence is more challenging to achieve. For
$V_{2}=3.5$, however, we can achieve a good convergence even at $U=0$,
as can be seen in Fig.$\,$\ref{fig:U-0_V2-3.5}. In this figure we
plot the exact and DMRG calculations of $\langle\delta n_{i}\rangle$
for $U=0$ and for different system sizes and see that only for $N=2008$
we start seeing more significant deviations.

\begin{figure}[h]
\centering{}\includegraphics[width=0.8\columnwidth]{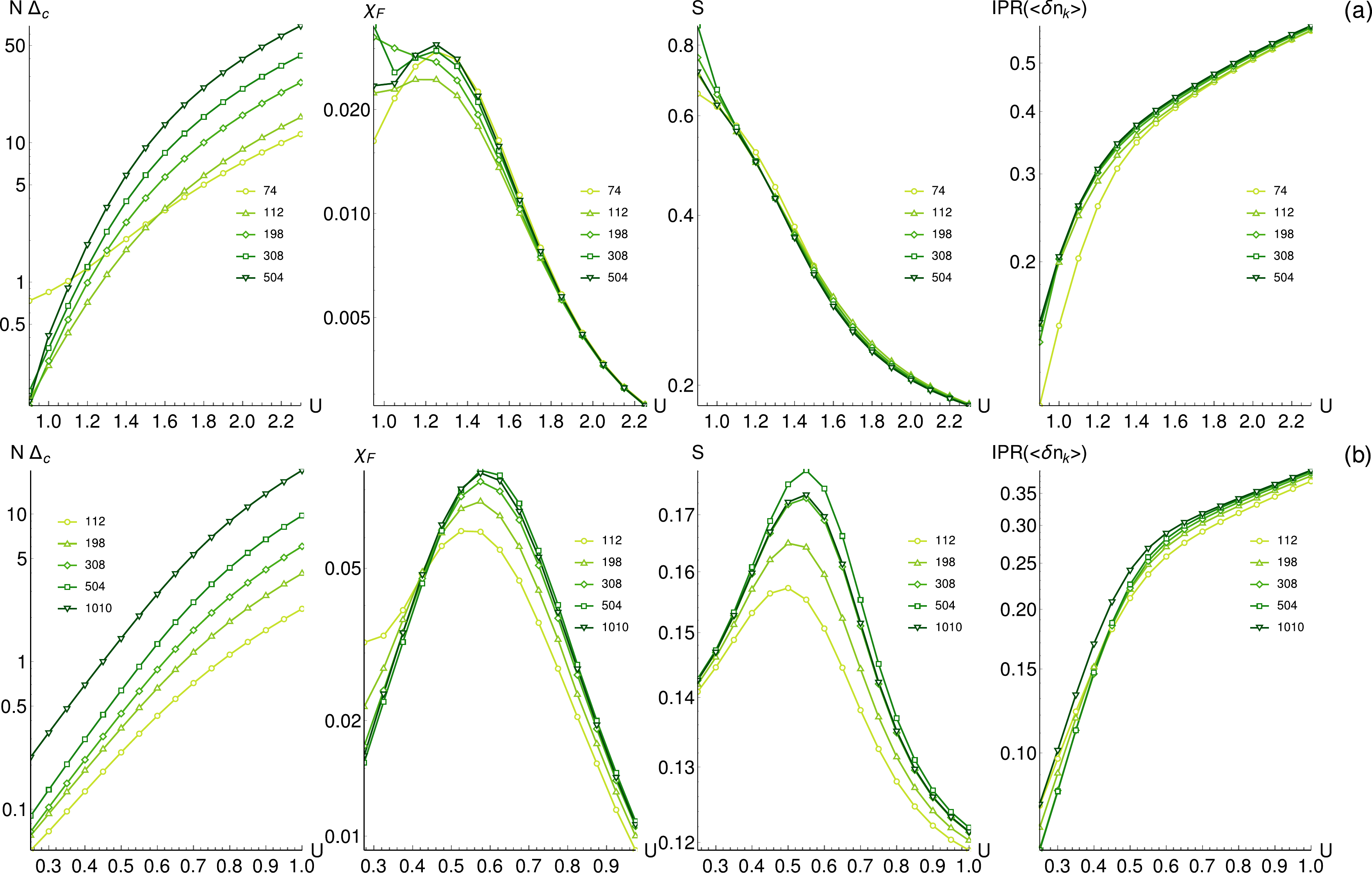}\caption{Additional results for $V_{2}=1.5,\phi=0.46$ (a) and $V_{2}=1,\phi=\pi/4$
(b). \label{fig:V2-1_1.5}}
\end{figure}

We also obtained results for other $V_{2}\geq1$. However, in this
case it was not possible to converge the results down to $U=0$ for
sufficiently large system sizes. In Fig.$\,$\ref{fig:V2-1_1.5} we
show results for $V_{2}=1$ and $V_{2}=1.5$ only for $\Delta_{c}\geq10^{-4}$,
for which there was a very good agreement between the DMRG and exact
results for $V_{2}=3.5$. In both cases, we see that $\chi_{F}$ has
a maximum that saturates with system size, again pointing to a crossover
instead of a true quantum phase transition. This is also corroborated
by the $S$ and ${\rm IPR}_{\kappa}(\langle\delta\bm{n}\rangle)$
results: $S$ also saturates with $N$ and ${\rm IPR}_{\kappa}(\langle\delta\bm{n}\rangle)$
does not decrease with $N$.

\section{Commensurate vs. incommensurate: Additional results}

\begin{figure}[h]
\centering{}\includegraphics[width=0.75\columnwidth]{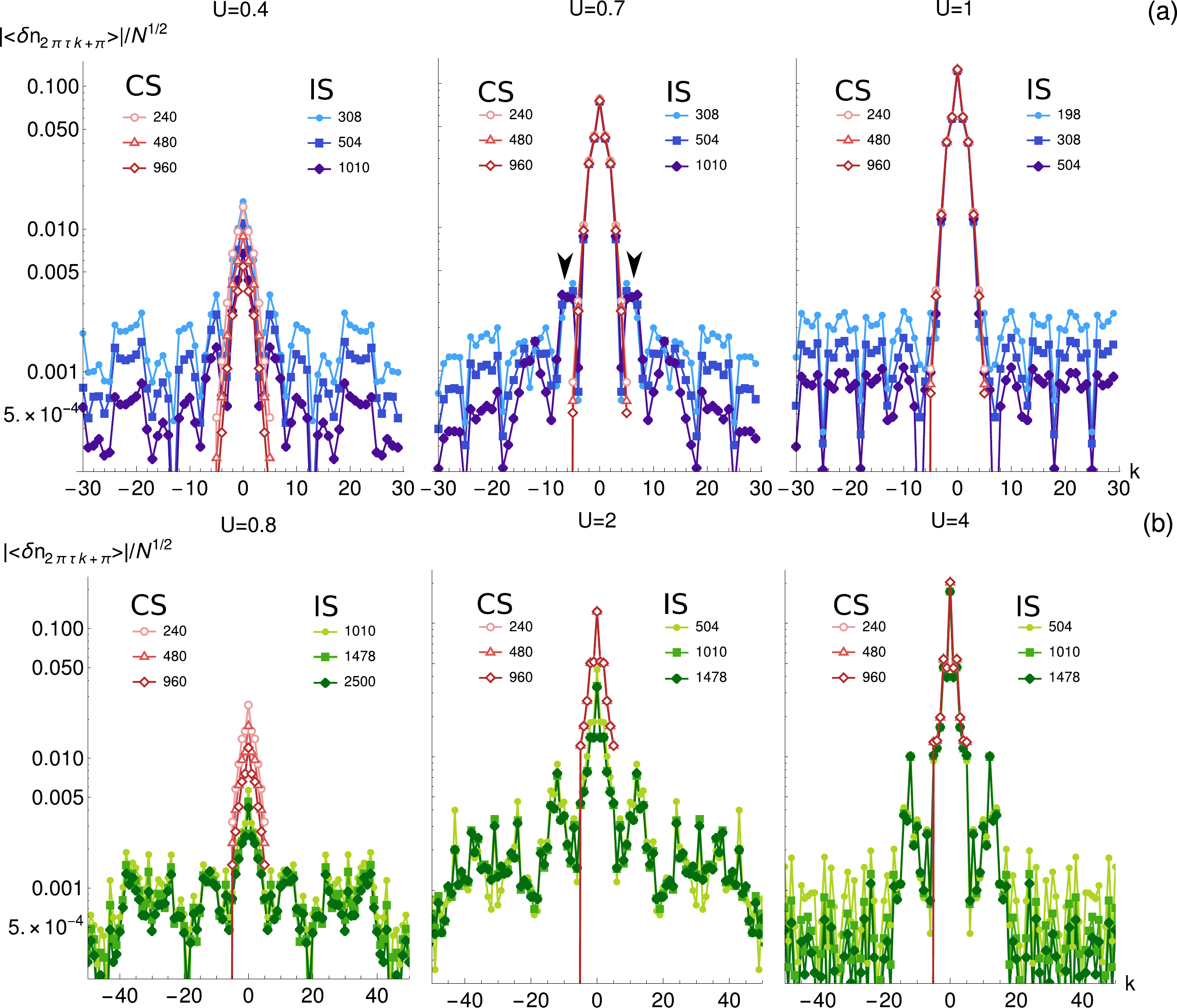}\caption{$|\langle\delta n_{\kappa_{k}=\pi+2\pi\tau k}\rangle|/\sqrt{N}$ for
CS and IS, for $V_{2}=0.975$ (a) and $V_{2}=3.5$ (b). For the CS
we used $\tau=\tau_{c}=7/12$ and $\phi=0$. In this case, there are
only $12$ inequivalent $\kappa_{k}$. For the IS we used $\phi=\pi/4$
in (a) and $\phi=1.8$ in (b). The arrows in (a) indicate ordered
peaks that exist for the IS and that are absent for the CS. For $V_{2}=3.5$,
in (b), we see that for smaller $U$ - left and middle panels - the
type of order is completely different for the IS and CS - in the former
case there are converged peaks up to very large $|k|$, while in the
latter $|k|$ is bounded and the ordered phase breaks down at sufficiently
small $U$. \label{fig:delta_nk_com_incom}}
\end{figure}

In this section we present some additional results with special focus
on the comparison between commensurate (CS) and incommensurate systems
(IS). For the CS, we take, as in the main text, a system with $\tau=7/12$,
having both the unit cell and the Moiré pattern with $12$ sites.
In the main text, we showed plots of $\langle\delta n_{\kappa}\rangle$
as a function of $\kappa$. In Fig.$\,$\ref{fig:delta_nk_com_incom}
we plot $|\langle\delta n_{\kappa_{k}}\rangle|$, with $\kappa_{k}=\pi+2\pi\tau k$,
as a function of $k$. This explicitly shows that CDW order is dominated
by wave vectors with smaller $|k|$, as mentioned in the main text.
Since for each approximant we have $\tau=p/N$ with $p$ and $N$
co-primes, this operation is just a reshuffling of the values $\kappa_{j}=2\pi j/N,\textrm{ }j=0,\cdots,N-1$.
For the CS, $\kappa_{j}$ only takes 12 different values, indicating
that the obtained CDW order has the same unit cell as the commensurate
system.

\begin{figure}[h]
\centering{}\includegraphics[width=0.6\columnwidth]{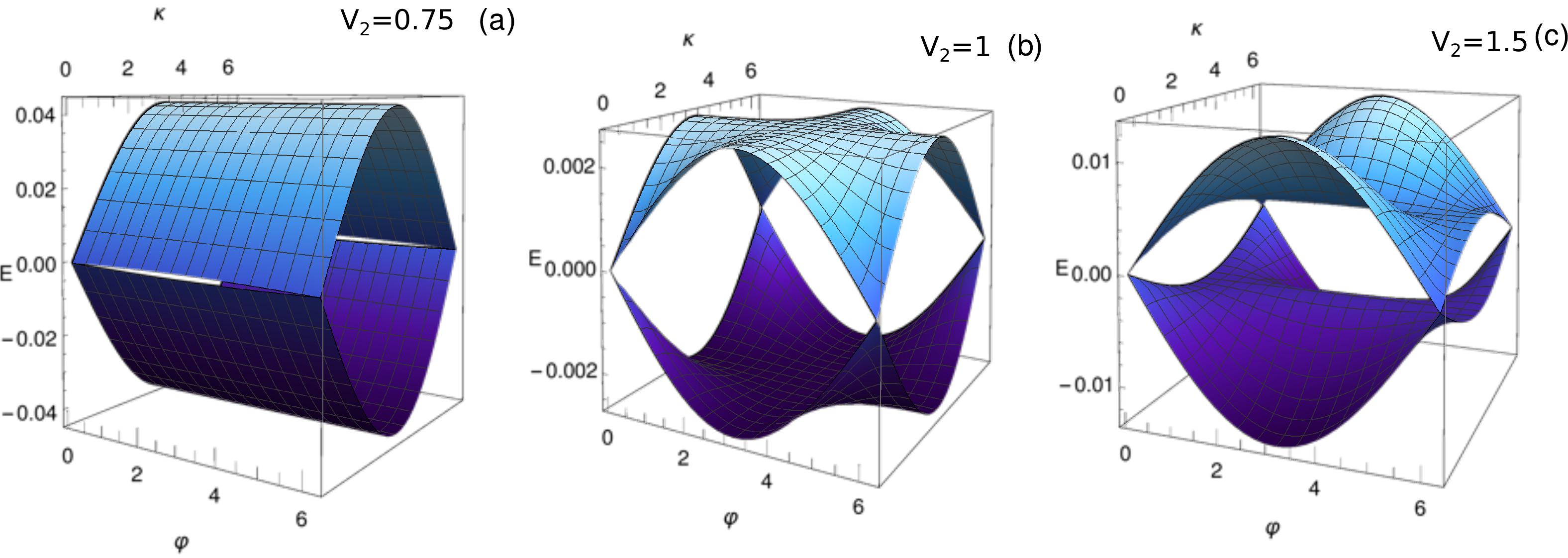}\caption{Dispersion of energy bands around $E=0$ for the CS with $\tau=7/12$
as function of the Bloch momentum $\kappa$ and $\varphi=12\phi$.
\label{fig:tau-7o12_phi_k_bands}}
\end{figure}

As we mentioned in the main text, depending on the choice of $\phi$,
an energy gap can open for the CS around $E=0$ in the non-interacting
limit. We show this in Fig.$\,$\ref{fig:tau-7o12_phi_k_bands}, where
we plot the energy bands around $E=0$ constituting the narrow-band
as a function of Bloch momentum $\kappa$ and $\varphi=12\phi$ (the
energy bands have a periodicity in $\phi$ of $\Delta\phi=\pi/6$).
In this figure, we can see that even for $V_{2}<1$, a small energy
gap can open. In fact, the system is only truly gapless for $\phi=\pi j/6,j\in\mathbb{Z}$.
For this reason, we chose $\phi=0$ for the results in the main text
so that the non-interacting limit is gapless both in the commensurate
and in the incommensurate cases.

\begin{figure}[h]
\centering{}\includegraphics[width=0.7\columnwidth]{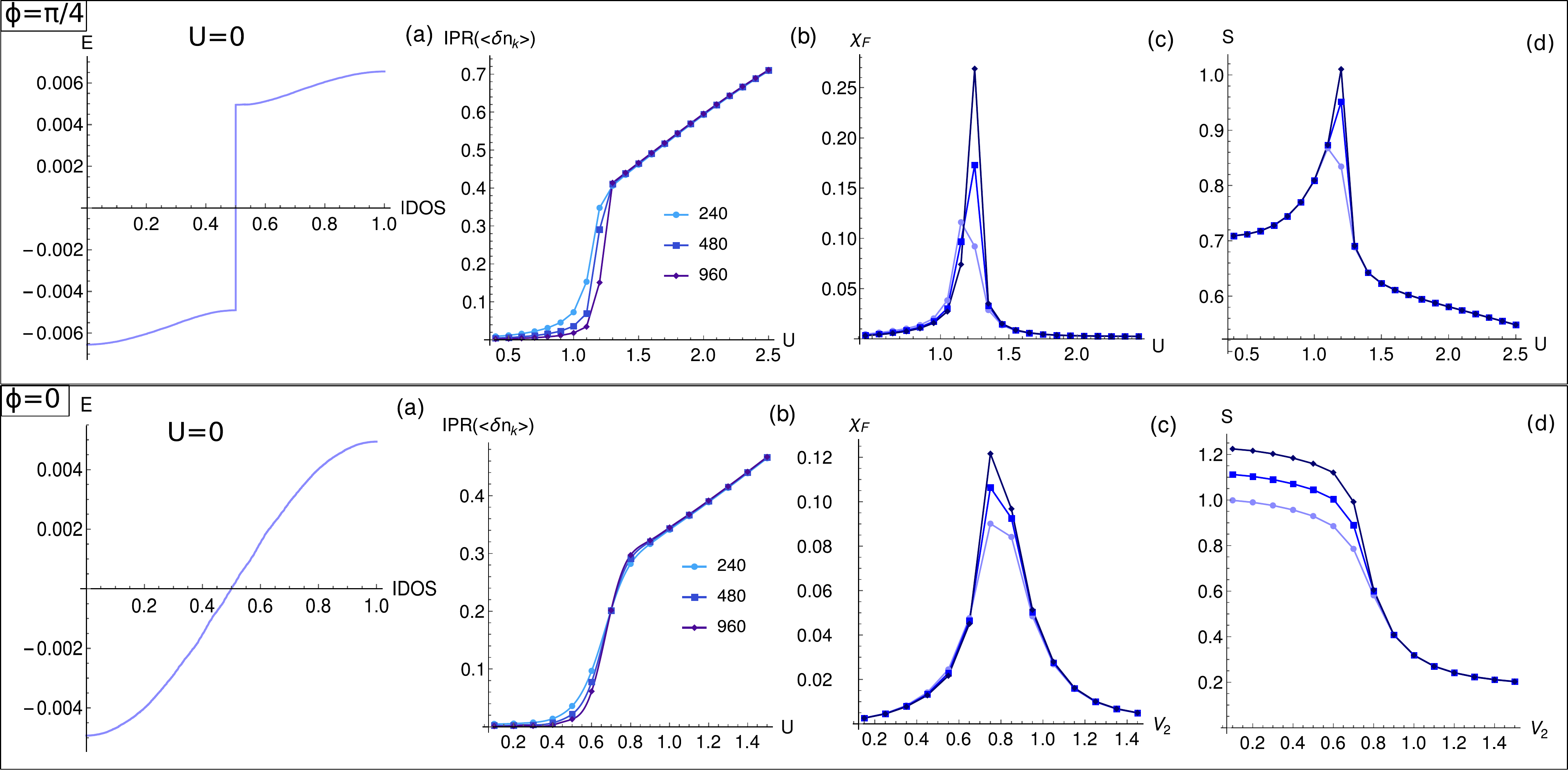}\caption{(a) IDOS for the CS with $\tau=7/12$, for $U=0,V_{2}=1.25$ and $\phi=\pi/4$,
gapped CS (top); $\phi=0$, gapless CS (bottom). (b-d): DMRG results
for the same parameters as in (a), for variable $U$, in the gapped
(top) and gapless (bottom) cases.\label{fig:V2-1.25_commensurate_transition}}
\end{figure}

One can also ask what happens if $\phi$ is chosen so that the CS
is gapped in the non-interacting limit. We observed that a phase transition
also occurs, but in this case between gapped disordered and gapped
CDW phases. Example results are shown in Fig.$\,$\ref{fig:V2-1.25_commensurate_transition}(top
panel) for $V_{2}=1.25$ and $\phi=\pi/4$ in which case the CS is
gapped for $U=0$; and, for comparison, for $V_{2}=1.25$ and $\phi=0$,
in which case the CS is gapless CS {[}Fig.$\,$\ref{fig:V2-1.25_commensurate_transition}(bottom
panel){]}. We can see that the critical $U$ is larger in the gapped
case. However, while the entanglement entropy at the middle bond,
$S$, increases with system size in the small-$U$ phase of the gapless
CS {[}Fig.$\,$\ref{fig:V2-1.25_commensurate_transition}(c-bottom){]}
- as it should in a LL phase - it saturates for the gapped CS, only
diverging at the critical point {[}Fig.$\,$\ref{fig:V2-1.25_commensurate_transition}(c-top){]}.
The gapped phase is therefore stable to finite interactions, up to
a considerably large critical $U$ at which the transition into the
commensurate CDW takes place.

\begin{figure}[h]
\centering{}\includegraphics[width=0.65\columnwidth]{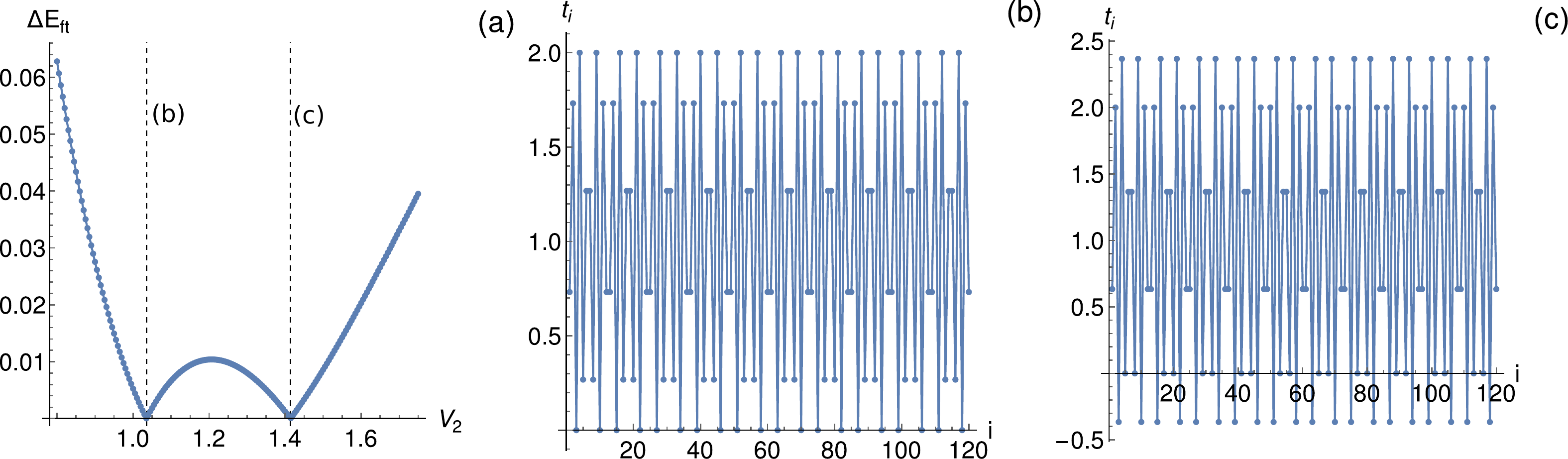}\caption{(a) Narrow-band width, $\Delta E_{{\rm ft}}$, in the non-interacting
limit, for the CS with for $\tau=7/12$ and $\phi=0$, as a function
of $V_{2}$. (b,c) Hopping configurations at the values of $V_{2}$
signaled by the dashed lines in (a). We can see that the narrow-band
becomes completely flat when hoppings vanish exactly. \label{fig:flatband_quench}}
\end{figure}

Finally, in the main text we have mentioned that for $V_{2}\geq1$
there are fine-tuned values of $V_{2}$ for which some hoppings vanish
and that in these cases, the narrow-band becomes completely flat.
These calculations are shown explicitly in Fig.$\,$\ref{fig:flatband_quench}.

\newpage

\section{Results with twisted-boundary-conditions}

In this section, we complement the results of the main text obtained
with OBC, by studying a system with twisted boundary conditions (TBC).
In this case, the system sizes that we can reach are significantly
smaller than for OBC. Nonetheless, the aim of this section is corroborate
the results obtained for OBC whenever possible. In the results that
follow, we take $\tau=\{27/46,43/74,65/112\}$, choosing the fraction's
denominator as the system size. TBC are applied in the Hamiltonian
as

\begin{equation}
\begin{aligned}H= & \sum_{j=0}^{N-2}t_{j}c_{j}^{\dagger}c_{j+1}+e^{i\kappa}t_{N-1}c_{N-1}^{\dagger}c_{0}+\textrm{h.c.}+U\sum_{j}n_{j}n_{j+1}\end{aligned}
,
\end{equation}
where $\kappa$ is the phase twist and again $t_{j}=-(1+V_{2}\cos[2\pi\tau(j+1/2)+\phi])$.
Below we present results for different configurations of $\phi$ and
$\kappa$.

\begin{figure}[h]
\centering{}\includegraphics[width=0.7\columnwidth]{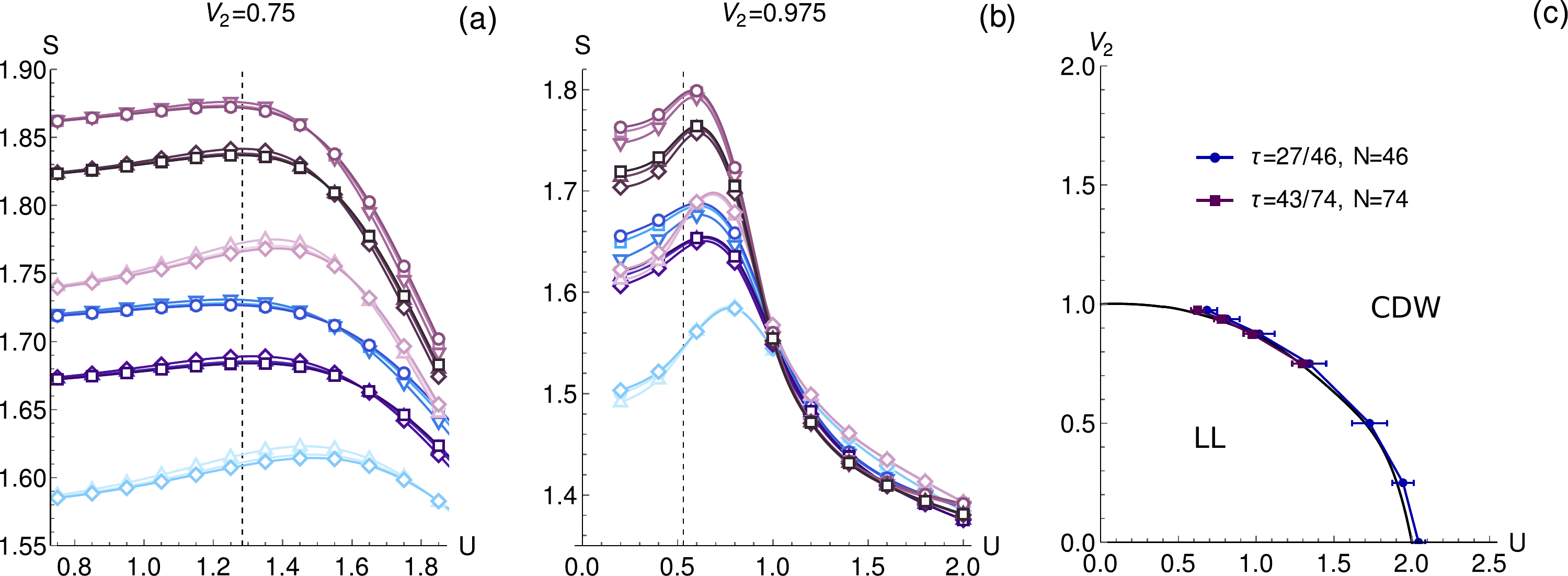}\caption{(a,b) Entanglement entropy at the middle bond, $S$, as a function
of $U$ for $N=46$ (blue) and $N=74$ (purple) and for $V_{2}=0.75$
(a) and $V_{2}=0.975$ (b). The dashed lines indicate the critical
point computed using open boundary conditions. The different tones
correspond to different configurations of $\phi$ and $\kappa$. We
used a grid of configurations $\Phi_{i+N_{\kappa}j}=(\phi_{i},\kappa_{j})=\Big(\frac{\pi}{2N}+\frac{2\pi i}{N_{\phi}},\frac{\pi}{2}+\frac{2\pi j}{N_{\kappa}}\Big),\textrm{ }i=0,\cdots,N_{\phi}-1;\textrm{ }j=0,\cdots,N_{\kappa}-1$,
where $N_{\phi}\times N_{\kappa}=3\times3$ is the total number of
configurations. The lighter (darker) color correspond to the $\Phi_{0}(\Phi_{8})$
configuration. (c) Maximizants of $S$, averaged over all the $9$
configurations along with the interpolation of the critical line obtained
using open boundary conditions (full black line). The standard deviation
is indicated by the error bars.\label{fig:S_TBC}}
\end{figure}

We start by presenting results for the entanglement entropy at the
middle bond, $S$, in Fig.$\,$\ref{fig:S_TBC}. The latter should
converge to a constant in the CDW phase, while increasing with system
size in the LL phase. There are still strong finite-size effects that
do not allow us to see the saturation of $S$ in the CDW phase close
to the critical point {[}Figs.$\,$\ref{fig:S_TBC}(a,b){]}. Nonetheless,
$S$ develops a maximum at a value of $U$ that does not depend very
significantly on $N$ and that very accurately matches the critical
point computed using OBC {[}see Fig.$\,$\ref{fig:S_TBC}(c){]}. For
$V_{2}\geq1$, finite-size effects become too severe to explore the
small $U$ regime.

Most importantly, in order to verify the type of order in the CDW
phase, we also compute the structure factor $S(q)$ defined as

\begin{equation}
S(q)=\frac{1}{N}\sum_{i=1}^{N}\sum_{j=1}^{N}[\langle n_{i}n_{j}\rangle-\langle n_{i}\rangle\langle n_{j}\rangle]e^{\textrm{i}q(i-j)}\label{eq:Sq}
\end{equation}

Order at $q=q_{0}$ is signaled by a diverging $S(q_{0})$ with $N$:
in the thermodynamic limit, $S(q_{0})/N$ should be finite in this
case, otherwise it should scale to zero. In Fig.$\,$\ref{fig:Sq},
we present results for $V_{2}=0.975$ and $V_{2}=1.75$ for $U$ within
the CDW phase. There we see that $S(q)$ only increases with $N$
at $q_{n}=\pi+2\pi\tau(n-1)$, the wave-vectors at which we have seen
the development of order using OBC. However, finite-size effects are
much stronger for TBC (due to the smaller available sizes) and it
becomes challenging to capture the divergence of $S(q_{n})$ for $n>4$.
Based on the results obtained for OBC, we conjecture that for large
enough system sizes, the quasi-fractal regime should be signaled by
the divergence of $S(q_{n})$ up to large order in $n$.

\paragraph{
\begin{figure}[h]
\protect\centering{}\protect\includegraphics[width=0.8\columnwidth]{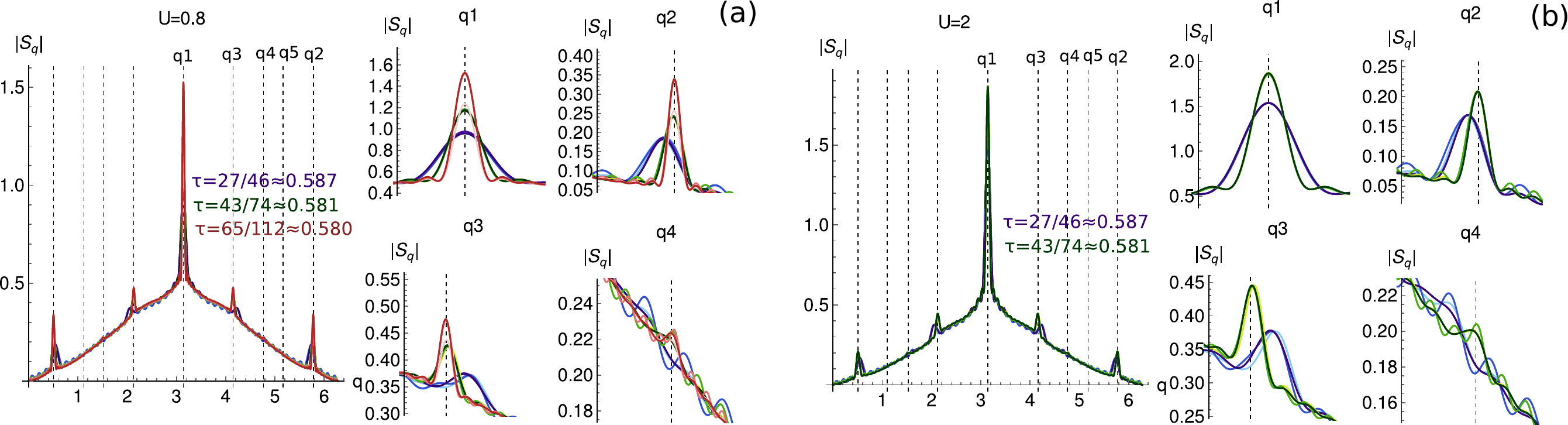}\protect\caption{Structure factor $S_{q}$, defined in Eq.$\,$\ref{eq:Sq} for $V_{2}=0.975,U=0.8$
(a) and $V_{2}=1.75,U=2$ (b). Curves with different lightness were
obtained for different configurations of $\phi$ and $\kappa$. The
dashed lines are located at $q_{n}=\pi+2\pi\tau(n-1)$. The right
smaller panels correspond to zoom-ins around $q_{n}$ up to $n=4$.
\label{fig:Sq}}
\protect
\end{figure}
}

\end{document}